\documentstyle[psfig]{mn2e}

\input{epsf}

\def\lsim{\mathrel{\hbox{\rlap{\hbox{\lower4pt\hbox{$\sim$}}}\hbox{$<$}}}}
\def\gsim{\mathrel{\hbox{\rlap{\hbox{\lower4pt\hbox{$\sim$}}}\hbox{$>$}}}}

\def\and {\rm {et al.} \rm} 

\begin{document}

\title[Clustering and descendants of MUSYC galaxies]
{Clustering and descendants of MUSYC galaxies at $z<1.5$}
\author[Padilla et al.]{
\parbox[t]{\textwidth}{
Padilla, N.D.$^1$, 
Christlein, D.$^2$,
Gawiser, E.$^3$,
Gonz\'alez, R.E.$^1$, 
Guaita, L.$^1$, 
Infante, L.$^1$.
}
\vspace*{6pt} \\ 
$^1$ Departamento de Astronom\'\i a y Astrof\'\i sica, Pontificia
     Universidad Cat\'olica de Chile, V. Mackenna 4860, Santiago 22, Chile.\\
$^2$ Max Planck Institut fur Astrophysik, Garching, Germany\\
$^3$ Astronomy Department,
     Rutgers University, USA.\\
}

\maketitle

\begin{abstract}
We measure the evolution of galaxy clustering out to a redshift of $z\simeq1.5$ using data from two MUSYC fields, the Extended Hubble Deep Field South (EHDF-S) and the Extended Chandra Deep Field South (ECDF-S).  We use photometric redshift information to calculate the projected-angular correlation function, $\omega(\sigma)$, from which we infer the projected correlation function $\Xi(\sigma)$.  We demonstrate that this technique delivers accurate measurements of clustering even when large redshift measurement errors affect the data.  To this aim we use two mock MUSYC fields extracted from a $\Lambda$CDM simulation populated with GALFORM semi-analytic galaxies which allow us to assess the degree of accuracy of our estimates of $\Xi(\sigma)$ and to identify and correct for systematic effects in our measurements.  We study the evolution of clustering for volume limited subsamples of galaxies selected using their photometric redshifts and rest-frame $r$-band absolute magnitudes.  We find that the real-space correlation length $r_0$ of bright galaxies, $M_r<-21$ (rest-frame) can be accurately recovered out to $z\simeq1.5$, particularly for ECDF-S given its near-infrared photometric coverage.  For these samples, the correlation length { is consistent with a constant value of $r_0=(2.6\pm0.3)$h$^{-1}$Mpc for the ECDF-S field, and $r_0=(3.0\pm0.4)$h$^{-1}$Mpc for the EHDF-S field from a median redshift $z_{med}=0.37$ to $z_{med}=1.15$.} There is mild evidence for a luminosity dependent clustering in both fields at the low redshift samples (up to $<z>=0.57$), where the correlation length is higher for brighter galaxies by up to $1Mpc/h$ between median rest-frame r-band absolute magnitudes of $\sim-18$ to $\sim -21.5$.  As a result of the photometric redshift measurement, each galaxy is assigned a best-fit template; we restrict to E and E$+20\%$Sbc types to construct subsamples of early type galaxies (ETGs).  These ETGs are separated into samples at different redshift intervals so that their passively evolved luminosities (to $z=0$) are comparable.  Our ETG samples show a strong increase in $r_0$ as the redshift increases, making it unlikely ($95\%$ level) that ETGs at median redshift $z_{med}=1.15$ are the direct progenitors of ETGs at $z_{med}=0.37$ with equivalent passively evolved luminosities.
\end{abstract}


\section{Introduction}
\label{sec:intro}

The evolution of galaxy clustering with redshift is a critical test
of the paradigm of galaxy and structure formation favoured today.  Large galaxy
redshift surveys such as the 2dF Galaxy Redshift Survey (2dFGRS, Colless et al. 2001) 
and the Sloan Digital Sky Survey (SDSS, York et al. 2000), have mapped the local
galaxy population.  Along with recent measurements from the cosmic
microwave background radiation by WMAP, 
they have constrained with unprecedented accuracy several cosmological parameters
(Dunkley et al., 2009, Spergel et al., 2007; S\'anchez et al., 2006).
The study of the evolution of galaxy clustering with redshift can allow us
to provide further tests to galaxy formation models by probing the evolution of the
amplitude of fluctuations in the quasi-linear to non-linear regime, which
only recently is becoming better understood by means of large numerical
simulations (Colberg et al. 2000; Sheth, Mo \& Tormen, 2001; Padilla \& Baugh, 2002; Slejak \& Warren, 2004).
On the other hand, studies on the evolution of clustering will also allow to probe
the evolution of the relation between galaxies and the underlying
dark matter, which is often parametrised by the bias parameter 
(e.g. Coil et al., 2004; Le F\`evre et al., 2005; Ouchi et al., 2004).

It is now becoming possible to perform deep galaxy redshift surveys, such
as the DEEP2 Galaxy redshift survey (Coil et al., 2004), the VIMOS VLT 
Deep Survey (VVDS, Le F\`evre et al. 2005), and the zCOSMOS redshift survey (Meneux et al., 2009)
which allow the study of the evolution of clustering with redshift.
Coil et al. (2004) obtained the evolution of the correlation length
using $2,220$ DEEP2 galaxies, Le F\`evre et al. (2005) using $\sim10,000$
VVDS galaxies, and Meneux et al. (2009) using a similar number of zCOSMOS galaxies.  
These studies, however, only reach redshifts of about $z=1$, mainly due
to the selection of galaxies by setting a lower limit on the photometric flux, which
preferentially selects galaxies at relatively moderate redshifts, and are affected by
cosmic variance due to the relatively small surveyed volumes.  

There are several 
techniques
that take advantage of features in the spectral energy
distribution (SED) to identify sets of galaxies at a particular redshift range.
These techniques 
use the Lyman Break in the SED at $z\sim 4$ and $z \sim 3$ (Ouchi
et al., 2004; Lee et al., 2006; Adelberger et al., 2005a, 2005b), or identify
galaxies with a Lyman-$\alpha$ line in emission at $z=2.1, 3.1, 4.5, 4.86$ (Guaita et al., 2010, 
Gronwall et al.,
2007, and Gawiser et al., 2007; Kovac et al., 2007; Ouchi et al., 2003, respectively),
with the H$\alpha$ line in emission (e.g. Sobral et al., 2010),
or use faint sources selected in the K-band (e.g. Quadri et al., 2007).  
Measurements of clustering from such samples, however,
are somewhat hindered by the relatively low numbers of galaxies. 
A recent measurement of the clustering of galaxies selected using the
Lyman$-\alpha$ emission (LAE galaxies) technique was performed by
Gawiser et al. (2007). 
They point out that different samples of galaxies at high redshift, such
as Lyman Break Galaxies and LAEs trace different underlying galaxy populations that can
be related via their clustering measurements with the aid of a theoretical framework.
Gawiser et al. first provide a thorough
comparison between different available clustering measurements from $z=0$ to
$z\simeq 5$, and then analyse whether samples at different redshifts can be considered to
be related in a parent/descendant relationship (see also Quadri et al., 2007, Francke et al., 2008, 
Guaita et al., 2010).  The latter is achieved
by comparing their measured clustering with expected
trends of the bias parameter with redshift extracted from simple analytic
approximations in a $\Lambda$CDM model.  This allows them to connect
the LAE population at $z=3.1$ to present-day $L^*$ galaxies.  

{ Another interesting approach is that presented by
Zheng, Coil \& Zehavi (2007, see also White et al., 2007, Brown et al., 2008, Wake et al., 2008), 
who fit the projected correlation functions
measured in SDSS and DEEP2 using the Halo Occupation Distribution model (HOD, see
for example Jing, Mo \& B\"orner, 1998, Peacock \& Smith, 2000, Cooray \& Sheth, 2002).
Using this powerful technique Zheng et al. are able to determine that the rate
of growth of the stellar mass is smaller in central than in satellite galaxies between redshifts $\sim1$ and $\sim0$,
and the fraction of stellar mass in satellites diminishes at high redshifts.  An interesting 
prospect is that of putting together theoretical models for the evolution of haloes with HODs and
applying the analysis to large sets of measurements of high-z galaxies.
}

Within this picture, intermediate redshift surveys such as DEEP2 and VVDS, which cover the range $0.3<z<1.0$, bridge
the gap between the low redshift surveys such as SDSS and
2dFGRS, and the samples of LAEs, LBGs and other photometrically selected galaxies at high redshifts.  
In this paper we will analyse the clustering of volume limited samples of MUSYC galaxies
\footnote{For full details on the survey see 
the MUSYC collaboration web-page http://www.astro.yale.edu/MUSYC.}, to complement
this intermediate redshift range, studying galaxy populations of 
various luminosities and redshifts out to $z=1.5$.  

The detailed evolution of the galaxy clustering
can also be used to study the assembly of passive galaxies.  This particular subject
has recently become the centrepoint of a discussion regarding galaxy evolution. It has
been claimed that high stellar mass, passive galaxies do not show an evolution
in their comoving space density.  This has been studied using the stellar mass and luminosity functions
in observations
(Cimatti et al., 2002, 2004; McCarthy et al., 2004;
Glazebrook et al. 2004; Daddi et al., 2005; Saracco et al., 2005;
P\'erez-Gonz\'alez et al., 2008; Bundy et al., 2006),
and has been interpreted as evidence that
their stellar content has already been assembled at high redshift,
ruling out the involvement of (even dry) mergers in the build up of their stellar mass.
Results from models of galaxy formation indicate that massive galaxies would continue
to acquire stellar mass at comparatively lower redshifts (e.g. De Lucia et al., 2006, Lagos, Cora \& Padilla, 
2008, Lagos, Padilla \& Cora, 2009).  
Measurements of clustering offer an alternative way to assess this problem 
which consists of connecting a population of galaxies at high redshift to 
a low redshift population, via the
expected evolution of clustering.  In this work we will test this approach using our MUSYC sample.

This paper is organised as follows.  We describe the MUSYC data in Section 2;
Section 3 describes the numerical simulation, the procedure followed to construct 
the mock catalogues,
and their role in our analysis.  The method used to infer the
clustering measurement is described in Section 4 and
thoroughly tested in Section 5, and
the results from the MUSYC galaxies are presented in
Section 6.  We discuss our results in Section 7  and
present our main conclusions in Section 8.

\begin{figure}
{\epsfxsize=8.5truecm 
\epsfbox[40 170 575 705]{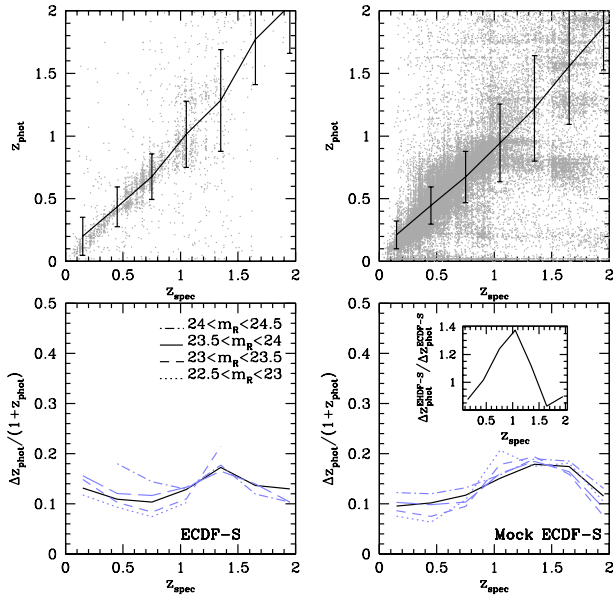}}
\caption{
Photo-z vs. Spectroscopic redshifts for the ECDF-S field (top-left) and
a mock catalogue (top-right).  The lower panels indicate the relative
error in redshift, as a function of the intrinsic galaxy luminosity (absolute
magnitude ranges are indicated in the key).
The inset on the lower right panel shows the ratio between the photo-z errors
obtained from mock catalogues with the photometry available in the EHDF-S and the ECDF-S (the former
does not include infra-red photometry).
}
\label{fig:zphoto}
\end{figure}

\begin{figure}
\begin{picture}(200,450)
\put(30,-10){\psfig{file=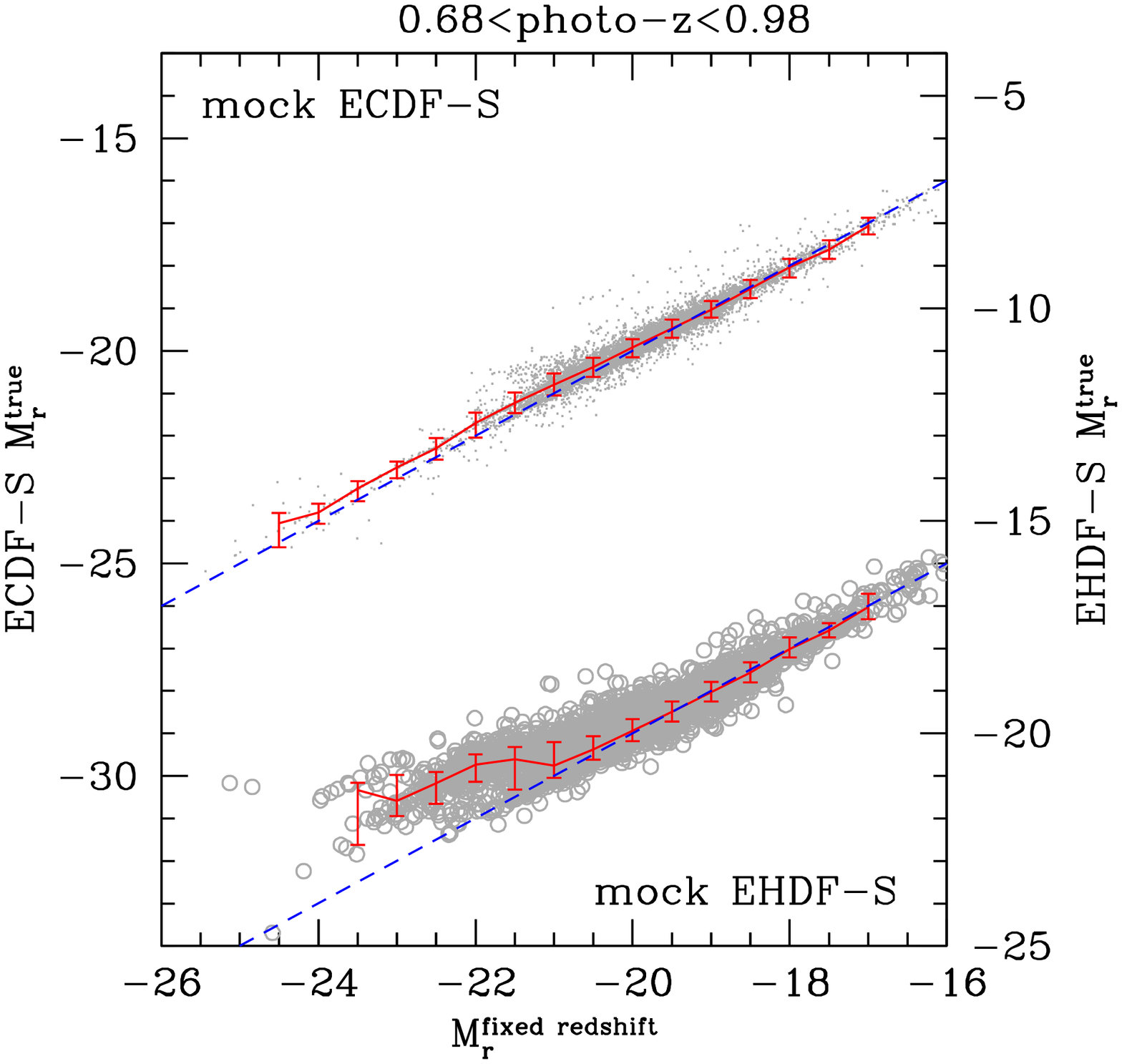,width=5.8cm}}
\put(30,140){\psfig{file=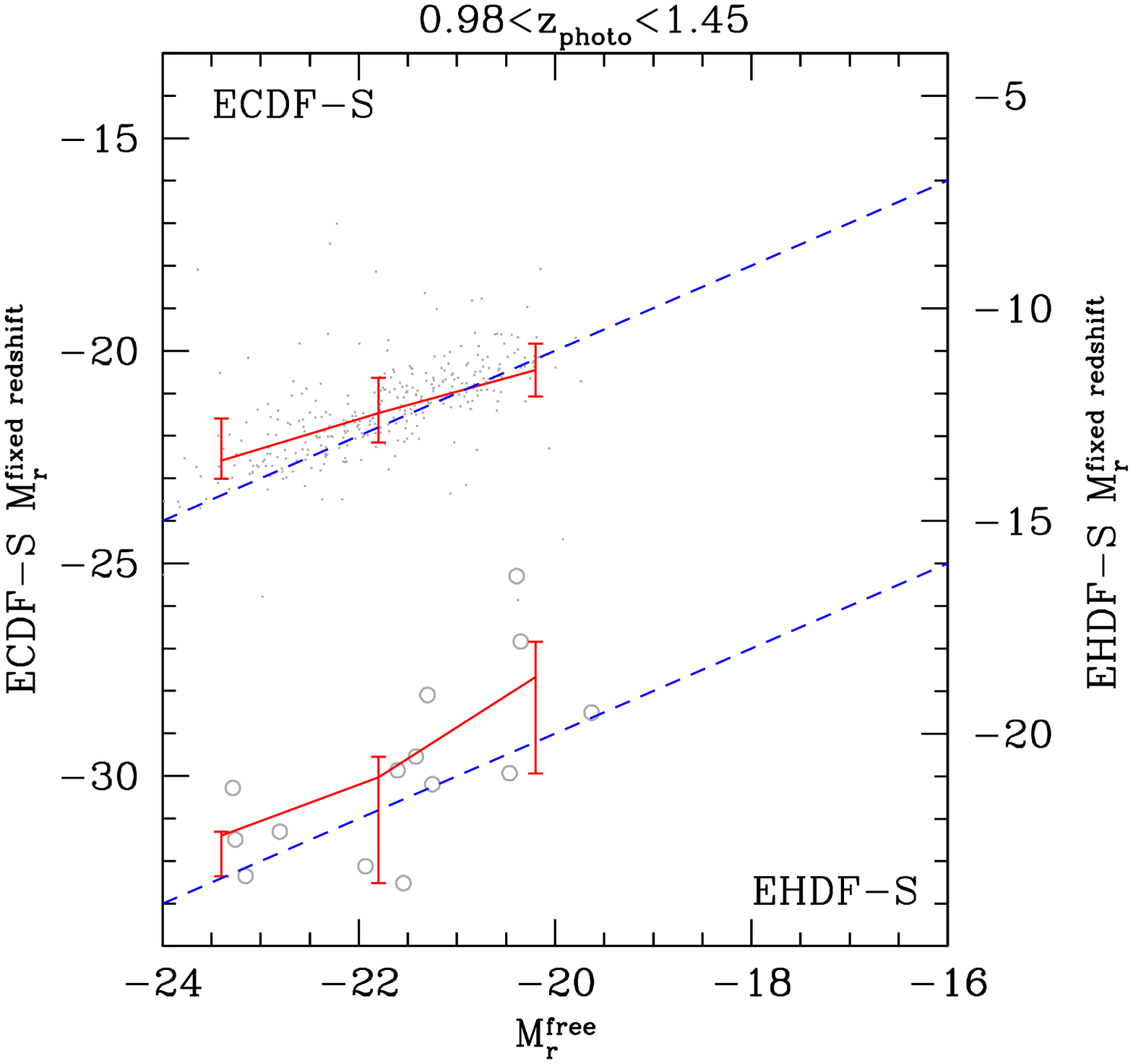,width=5.8cm}}
\put(30,290){\psfig{file=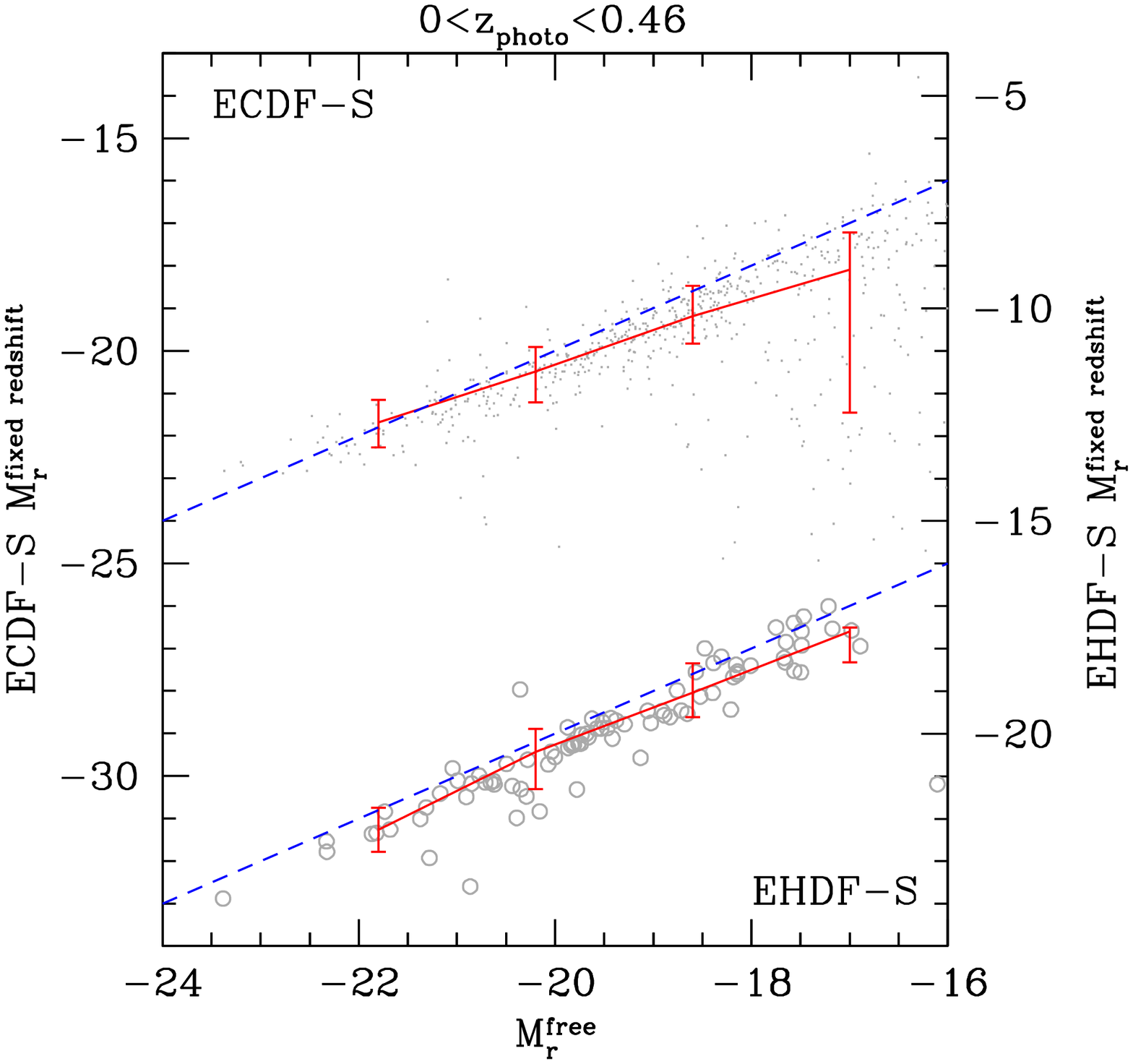,width=5.8cm}}
\end{picture}
\caption{
Top and Middle panels: Recovered ($M_{\rm free}$) vs. true ($M_{\rm fixed redshift}$)
absolute magnitudes in the $r$ band for galaxies in the ECDF-S (dots, upper
line) and EHDF-S (open circles, lower line) fields.  
Each panel shows a different redshift slice,
indicated on the top.  
The left-hand (right-hand) 
y-axis shows the $M_r$ scale for the ECDF-S (EHDF-S).  The lines with errorbars show the
median and $10$ and $90$ percentiles, respectively.  
The dashed lines show the one-to-one relation.
Bottom panel: recovery of the true rest-frame r-band absolute magnitude in the Bower
et al. mocks, when fixing the redshift at its spectroscopic value.  Lines and symbols
are as in the top two panels.
}
\label{fig:rmag}
\end{figure}

\section{MUSYC}
\label{sec:data}
The Multi-wavelength Survey by Yale-Chile (MUSYC) comprises
$1.2$ square degrees of sky.  
The full survey covers four fields
chosen to leverage existing data and to enable flexible scheduling of observing
time during the year.  Each field is imaged from the ground in the optical and near-infrared.  
Follow-up spectroscopy is done mostly with multi-object
spectrographs (VIMOS and IMACS).  The survey fields will be a natural choice for
future observations with ALMA.
The data used in this paper corresponds to two MUSYC fields with the most
complete photometric and spectroscopic coverage, the EHDF-S and ECDF-S fields. 
These fields include imaging in the $U,B,V,R$ filters
to AB depths of $26.5$ plus additional shallower imaging in $I$ and $z$.
The $ECDF-S$ field also includes $K,J$ imaging to a depth of $22.5$ AB.
Both fields include 
extensive follow-up spectroscopy of specific colour-selected galaxy samples.

In MUSYC, galaxies are defined as sources identified using SExtractor
(Bertin \& Arnouts, 1996) { on the co-added $BVR$ images}, with SExtractor stellarity
parameter $c<0.8$ (Huber, 2002
\footnote{Huber (2002) estimate a SExtractor stellarity parameter
$c=0.8$ as the most adequate value for separating stars ($c>0.8$)
and galaxies.  Our results do not change significantly when
varying the stellarity threshold from $0.75$ to $0.85$.}).  
The total number of galaxies in MUSYC is { $\sim280,000$ at $m_{BVR}<27.1$}; after restricting
our samples, this number decreases to $\sim50,000$ galaxies.

\begin{figure}
{\epsfxsize=8.truecm 
\epsfbox[40 170 575 705]{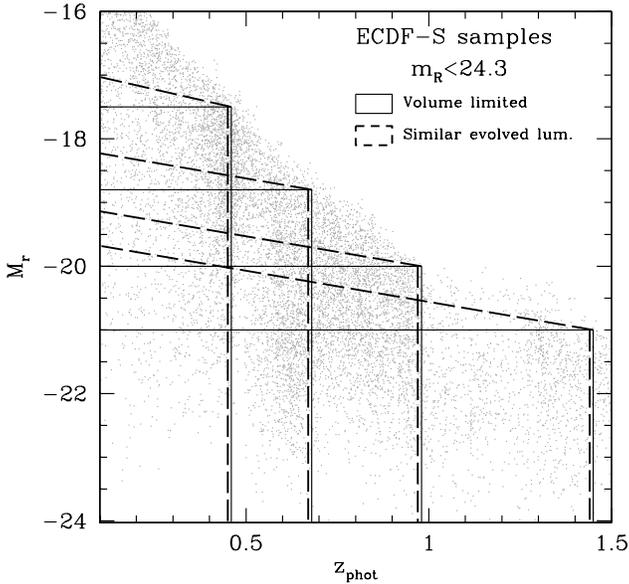}}
\caption{
Construction of centre galaxy samples for our
cross-correlation measurements.  The points represent galaxies in the
ECDF-S field with apparent magnitudes below the completeness limit,
$m_R<24.3$.  The vertical lines show the adopted limits in redshift.  Horizontal
lines show different limits in rest-frame absolute magnitudes $M_r$ used as
further constraints in the construction of galaxy samples.
The long-dashed lines show the limiting luminosities adopted for samples of
early-type galaxies selected so as to have similar passively evolved luminosities
at different redshifts.
}
\label{fig:hd}
\end{figure}

Photometric redshifts were estimated using a least squares frequentist best-fit 
(e.g. Bayarri \& Berger, 2004) to
the observed photometry of MUSYC galaxies.
{ The templates in the synthetic set adopted by HYPERZ (Bolzonella, Miralles \& Pell\'o, 2000),
allowed to evolve with time, are used to represent the redshift/spectral type mix in
MUSYC.  Once a photometric redshift and a best fitting evolving template are assigned to a galaxy, the 
latter is compared to a collection of $11$ templates selected as in Christlein et al. (2009).
The Christlein et al. templates correspond to different $z=0$ galaxy morphologies
from early types (E, E$+20\%$Sbc) to late type Irr galaxies, and allow us to 
define a galaxy morphology from the best fitting evolving template.} 
When calculating the  redshift probabilities, we used all
available MUSYC flux measurements from all the available bands.
As can be seen in the top-left panel of Figure \ref{fig:zphoto},
the resulting photometric redshifts out to $z_{spec}\simeq 1.5$ 
agree to roughly $30\%$ with the spectroscopic redshifts.  This analysis
arises from a sample of $3193$ galaxies with available spectroscopic
redshift measurements in the ECDF-S from available MUSYC spectroscopy and the 
NED database.  The lower-left panel shows in different line types different
limits in R-band apparent magnitudes, and indicates that photometric
redshift (photo-z) errors change only slightly with galaxy luminosity.  The lack of
severe degeneracies is also clear in our photo-z estimates.  This comparison indicates
that the error in the estimate of photometric redshifts is
$\Delta z=0.09(1+z)$ after removing $5-\sigma$ outliers {(The normalised median absolute
deviation, NMAD, calculated according to Hoaglin et al., 1983, is 0.065)}.  The remotion of outliers
diminishes the sample of ECDF-S galaxies by a $10\%$.  
We do not use galaxies fainter than $m_R=24.3$ since at these apparent magnitudes
the quality of the photometric redshift estimation starts to degrade.  The dots
in the figure represent each galaxy in the ECDF-S
down to this magnitude limit.  
{ For EHDF-S 
the total number of spectroscopic redshifts available amounts to $405$ (much less than
for ECDF-S).  The comparison between spectroscopic and photometric redshifts shows
that photo-z errors are consistent with those for
ECDF-S, even when there is no near-infrared photometry for this field.  The remotion
of the $5-\sigma$ outliers diminishes the EHDF-S sample by a $16\%$.
As will be shown in Section 3, the analysis of mock catalogues with the photometry
available in the EHDF-S and ECDF-S suggests that the photo-z errors in the former
will be up to a factor $1.4$ higher (see the inset in the lower-right panel of the figure),
as a result of the lack of infra-red photometry in this field.  This also increases
the amount of catastrophic errors in EHDF-S; the mock catalogues will allow us to assess
the degree out to which this will affect the EHDF-S results.}

{ The upper and middle panels of Figure \ref{fig:rmag} show the comparison between the rest-frame 
r-band absolute magnitudes ($M_r$) estimated using the spectroscopic and photometric redshifts 
of each galaxy, for the EHDF-S and ECDF-S fields, for two different redshift slices (indicated
in each panel).  
In all cases, the photo-z estimates of $M_r$ show offsets of up to $\simeq 1$ magnitude from
the spectroscopic ones.  In the case of EHDF-S, at $z_{\rm photo}>0.98$, this
difference may be higher, of up to $\sim 2$ mags, but
the low number of spectroscopic redshifts available does not allow to reach a firm
conclusion on this regard; however, the analysis of mock catalogues will confirm that the
recovery of $M_r$ is expected to be significantly better in ECDF-S (see the lower panel).  
}

In our clustering measurements we will use two types of galaxy samples.  Centre
samples around which the clustering will be measured, and tracer samples that will
be used to trace the structure around each centre galaxy.
We restrict centre galaxies to the redshift range $0.1<z_{\rm photo}<1.45$
and divide the data in four ``centre" subsamples, defined by 
photo-z ranges limited by $0.1-0.46$, $0.46-0.68$, $0.68-0.98$ and $0.98-1.45$;
these limits are set so as to probe a wide redshift range.
The resulting median photometric redshifts are $z=0.38,0.46,0.81$ and $1.14$, and
the sampled volumes are approximately
$80,000$h$^{-3}$Mpc$^3$,
$129,000$h$^{-3}$Mpc$^3$,
$262,000$h$^{-3}$Mpc$^3$, and
$550,000$h$^{-3}$Mpc$^3$ for the
lower to higher redshift samples, respectively.
The aim in dividing the sample in different bins of redshift
is to be able to detect a variation in
the clustering of galaxies with redshift.  
Additionally, we apply rest-frame r-band absolute magnitude cuts
to construct volume limited samples; these are illustrated by the solid lines in 
Figure \ref{fig:hd}.
Since we will use a cross-correlation technique, 
we define samples of tracer galaxies, which are galaxies selected
using the same absolute magnitude cuts, but allowing a further $\delta z=0.1$ to
each side of the corresponding redshift interval { (with the restriction
of maximum redshift $z=1.5$)}.  This has the advantage of
improving the statistical signal while only inducing small correlations
between results from different redshift slices since the centre samples do not overlap
{ (we do expect correlations arising from the photo-z errors which will contaminate
neighbour redshift bins, and also from fluctuations of large-scale modes
straddling the centre samples).}

\begin{figure}
\begin{picture}(250,127)
\put(5,-15){\psfig{file=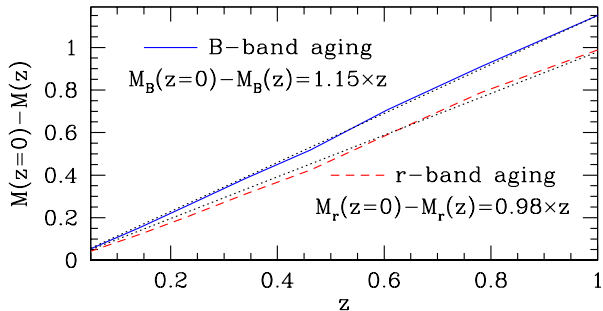,width=8.5cm}}
\end{picture}
\caption{
Variation in the absolute magnitude of a SSP ($3.5$Gyr old at $z=1$) as a function of redshift,
in the B- and r-band (solid and dashed lines, respectively).  The dotted lines correspond
to the fits indicated in the figure key.  The upper line shows the recipe adopted by Cimatti et al.
for the B-band, the lower dotted line shows the best fit to the evolution of the SSP in the r-band.
}
\label{fig:evo}
\end{figure}

We also construct samples of early-type galaxies with similar
passively evolved luminosities.  {
In order to select ETGs, we find the $z=0$ template that best matches the
evolving template that was originally fit to the galaxy in the photo-z calculation.
This effectively selects ETGs with similar colours at different redshifts.
In order to select similar passively evolved luminosities,} we use the empirical passive evolution recipe adopted
by Cimatti et al. (2006) for the B-band, where 
\begin{equation}
M_B(z=0)=M^{rf}_B(z)+1.15\times z,
\label{evoB}
\end{equation}
where $M_B(z=0)$ is the absolute magnitude of a galaxy passively evolved to $z=0$; at
redshift $z$ the galaxy has a rest-frame absolute magnitude $M^{rf}_B$.
This recipe is derived empirically from the evolution of the Fundamental Plane for 
massive early-type galaxies (di Serego Alighieri et al., 2005).  
As our samples are characterised by r-band magnitudes we find the corresponding
evolution recipe in the r-band using models for the spectra
of early-type galaxies (ETGs) approximated by single stellar populations (SSP).
We first look for the SSP that best matches the evolution
recipe from Eq. \ref{evoB} using spectral energy distributions (SED)
constructed using
Bruzual \& Charlot (2003, BC03) stellar synthesis algorithm and the Padova 2000 evolutionary
tracks with a Salpeter initial mass function and a metallicity $[Fe/H]=0.3$ -
this particular metallicity is adequate for high mass ETGs (e.g. Gallazzi et al.,
2006).
We find that a SSP which at $z=1$ is $3.5$Gyr old provides the best match
to the recipe of Eq. \ref{evoB} (consistent with studies by Bell et al., 2004).  This
can be seen in Figure \ref{fig:evo} which shows the evolution
of the B-band magnitude for this SED along with the recipe from
Eq. \ref{evoB}.  The figure also shows the evolution of the r-band magnitude for the same SEDs 
along with the best fit to the r-band evolution which corresponds to
\begin{equation}
M_r(z=0)=M_r^{rf}+0.98 \times z, 
\label{evor}
\end{equation}
where $M_r^{rf}$
is the rest-frame luminosity at redshift $z$.
We use the relation from Eq. \ref{evor} to construct 
samples with comparable evolved $z=0$ r-band luminosities
delimited by 
$M_r(z=0)=-16.85,-18.2,-19$, and $-19.6$.  
We notice that the evolution recipe adopted here is particularly adequate for massive ETGs
(di Serego Alighieri et al., 2005); 
even though we apply it to faint galaxies as well, we will only make comparisons between different
redshift bins for the brightest samples.
We apply the same redshift
cuts used in the construction of the volume limited samples (see above).  
The resulting samples are illustrated
in Figure \ref{fig:hd} by the
dashed lines (the vertical lines have been slightly displaced to improve clarity);{ the dependence
of $M_r$ on redshift shown by these lines correspond to the evolution of the ETG SED
from Eq. \ref{evor}, and as can be seen, the ETG samples selected this way should be complete
since the dependence of the selection effect is stronger than the modeled passive evolution.  
Given that the detection of sources is done using optical filters (B,V and R),
it could be possible that this will act against a complete selection of ETGs
as proposed here.  However, the detection limit in the combined optical photometry
is $2.8$ magnitudes deeper
in flux than the sample selection cut.  Still, in order to check whether this is affecting our 
results, we will also use the available imaging in the $K$ band to produce an alternative sample
of ETGs for the ECDF-S (this is not available for EHDF-S); the equivalent magnitude limit that
produces a sample of ETGs comparable the one obtained in the R-band is $m_{K,lim}=22.5$.}

{ We note that our samples of ETGs will consist of read-and-dead galaxies at all the redshift
intervals chosen.  As we will show in Section 6, our approach is consistent with selecting
galaxies in the red branch of the bimodal colour distribution, that is, in the red sequence.
The latter has been widely used to study ETG properties and clustering evolution, as in Bell et al. (2004),
Faber et al. (2006), and Coil et al. (2008).  Another possibility for the selection of ETG samples
in a direct descendant line at different redshifts is to let the colour of
the galaxy templates evolve, but in this case we find that
the selected galaxies either include blue cloud objects in our
higher redshift samples or too few objects at low redshifts, mainly due to the small solid angle of
the MUSYC survey.  This can be taken as a possible indication that the samples of ETGs constructed
by selecting galaxies in the red sequence with similar $z=0$ passively evolved luminosities, may not
constitute the parents and descendants of one another.  However, small amounts of star-formation
could easily change the colours of ETGs, making this comparison difficult.  In order to overcome
this problem, from this point on 
we will only analise the possible parent/descendant relationship between 
ETG MUSYC samples at a given redshift and SDSS ETG samples at $z=0$.}

As over the individual redshift ranges probed by our analysis
the change in clustering amplitude is expected to be roughly linear with
redshift,
our clustering measurements will be quoted at 
the median redshift of the galaxy subsample.

\section{Mock catalogues}

Due to the large uncertainties
in the determination of photometric redshifts it is necessary to assess the effects arising from
possible sources of systematics in our procedure.  In order to do this,
we use two mock catalogues extracted from a 
$\Lambda$CDM numerical simulation
populated with GALFORM (versions corresponding to { Baugh et al., 2005, and Bower et al., 2006})
semi-analytic galaxies, kindly provided by the Durham group.
As the underlying clustering and its evolution in the simulation are known, the
results from the mock catalogues can be used to find
systematic biases in our estimates, and to devise a method to account for them using
only data available in the observational sample.
The clustering of galaxies in the
simulation mimics reasonably well the clustering of real galaxies, which
makes the use of these mocks appropriate to test
our method.

{ We construct two of mock catalogues, one of them
using a single $z=0$ simulation output (and therefore
with a constant clustering with redshift) with Baugh et al. (2005) galaxies.  The other mock
is constructed using simulation
outputs at different redshifts (i.e. an evolving lightcone) containing Bower et al. (2006) galaxies; 
this mock catalogue includes the evolution of
the galaxy population and their clustering with redshift 
as results from the adopted cosmology and the assumptions 
in the semi-analytic galaxy formation model of Bower et al.
In both cases, subsamples of different rest-frame luminosities show different
clustering strengths.  The ability of our measurement method to detect
these variations will allow a study of the reliability of any clustering
dependence on luminosity and redshift found in the real data.}

There are a total of $10^9$ particles in this simulation,
the box side measures $1000$h$^{-1}$Mpc a side, the
matter density parameter corresponds to $\Omega_m=0.25$,
the value of the dark energy density parameter is $\Omega_{\Lambda}=0.75$, 
the Hubble constant, $H=100h$kms$^{-1}$Mpc$^{-1}$, with $h=0.7$,
and the primordial power spectrum slope, $n_s=0.97$.  The
present day amplitude of fluctuations in spheres of
$8$h$^{-1}$Mpc is set to $\sigma_8=0.8$.  This particular
cosmology is in line with recent cosmic microwave background
anisotropy and large scale structure 
measurements (WMAP team, Dunkley et al., 2009, Spergel et al. 2007, S\'anchez et al., 2006).  
We have adopted this cosmology for all the calculations performed throughout this paper.

The mock catalogues are constructed by selecting a suitable direction in the
simulation cube and following it out to $z=3$ replicating the simulation
as many times as needed ($4$ in total, { including outputs at
different redshifts in the Bower et al. case)}.  The direction is selected such
that the structures sampled by the cone are not repeated.
The estimates of uncertainties
will be obtained using the jacknife technique applied to each mock field
individually.  This technique has been shown to
provide uncertainties comparable to the scatter in clustering results
from large numbers of individual mock catalogues (see for instance, Padilla et
al., 2001).

The process of assigning galaxies to the mock catalogues consists of checking
that the angular position of the galaxy falls within the MUSYC angular
mask, and by placing the same apparent and absolute magnitude limit cuts as in 
the MUSYC data defined in Section \ref{sec:data}.
These magnitude limit cuts imprint a radial selection function in 
the mock catalogues which is qualitatively similar to that of the real
data (obtained using photo-zs).

We apply the least squares frequentist best-fit method 
to assign photometric redshifts to mock 
galaxies, replicating the process followed for the observational data { as well
as the available photometry  (i.e. different mocks for EHDF-S and ECDF-S galaxies). }
A comparison between underlying (spectroscopic) and photometrically derived
redshifts { for a Bower et al. ECDF-S mock} is shown in the right 
panels of Figure \ref{fig:zphoto}; the
lower right panel shows that the photo-z redshift uncertainties are comparable
to those present in the real data (left panels).  
Additionally, the top-right panel shows
a similar pattern as the MUSYC galaxies, with structures in the scatter plot 
situated at particular values of
spectroscopic redshift reflecting the large-scale structure, and at certain values 
of photometric redshift due to attractors in the photometric fitting solution.
This comparison further ensures us we have a proper tool to determine
the statistical and systematic errors in the measured clustering
amplitude arising from realistic photo-z errors.
{ We calculate the photo-z errors for a Bower et al. EHDF-S mock, and perform the
ratio between these errors and those obtained for the ECDF-S mock.  This ratio
is shown in the inset of the lower-right panel of Figure \ref{fig:zphoto} where it can be
seen that the lack of near infrared data in EHDF-S results in photo-z errors of up
to a factor $1.4$ larger than in the ECDF-S mock.}

{ Using the mocks, we also test whether the adopted template set allows us to recover the 
true rest-frame absolute magnitude in the r-band from the available photometry.  To do this
we apply the template fitting procedure fixing the redshift at its spectroscopic (true) value.
The comparison
indicates that for the ECDF-S photometry
the recovery is very precise, with systematic offsets lower than $0.2$ magnitudes.
The EHDF-S shows a good recovery for galaxies of all luminosities at $z<0.45$; at higher
redshifts this is true only for faint galaxies with $M_r>-21.5$.  Brighter galaxies
show inferred magnitudes systematically brighter by $\sim 1.5$ mags,
as can be seen in the example shown in the bottom panel 
of Figure \ref{fig:rmag}.  Such an offset was expected to some degree since
the available photometry in EHDF-S, UBVRIz, only allows to obtain an extrapolated
rest-frame r-band magnitude at redshifts $z>0.5$.
We will bear in mind this possible systematic
offset when analysing the EHDF-S results at this redshift range.}

\section{Method}
\label{sec:method}

Our aim is to obtain a reliable measurement of the real-space correlation length $r_0$,
the separation at which the 3D spatial correlation function satisfies $\xi(r_0)=1$.
In the following description we will use the term ``redshift" to refer to photo-zs in the case of MUSYC
data, and to refer to either photometric or spectroscopic redshifts in the case of the 
mock catalogues { (in this case, spectroscopic redshifts correspond to the true galaxy redshifts)}; 
when analysing the latter, spectroscopic redshifts will be used to infer the true underlying clustering 
present in the mock samples.  We will apply the following steps both to real and mock data:
\begin{itemize}
\item[(i)]
Measure the projected-angular cross-correlation function $\omega(\sigma)$ as a function of the
comoving projected separation, $\sigma$.  { When calculating this correlation function
one assumes
that all tracers (usually with no distance information) lie at the known distance of the centre galaxy,
given by its redshift (spectroscopic or photometric).
In our case 
this approach keeps the effect of distance measurement errors to a minimum by only 
using photometric redshifts to estimate comoving distances to the centre galaxies, 
and to restrict the range of redshifts of tracers (i.e. we never calculate the relative distance
between galaxies in the radial direction);
this is the main aim behind the choice of this cross-correlation function.}

Centre samples comprise galaxies selected by applying the cuts in redshift (spectroscopic or
photometric) and absolute magnitude (evaluated at the redshift of each individual galaxy)
defined in Section \ref{sec:data}.  The tracer samples are characterised by the
same cuts in rest-frame absolute magnitude (calculated at the redshift of each individual
galaxy) and by redshifts
$z_{min}-\delta z<z<{\rm min}(1.5,z_{max}+\delta z$),
where $z_{min}$ and $z_{max}$ are the limits of the centre sample, and $\delta z=0.1$.
The wider redshift range allowed for tracers results in an increase of the number of pairs
around centre galaxies near $z_{min}$ and $z_{max}$.

The estimator applied in this case is $$\omega(\sigma)=D_C D_T/D_C R-1,$$ where
$D_C D_T$ and $D_C R$ are counts of pairs of centre vs. tracer, and centre vs.
random points, respectively.  Random points are extracted from random catalogues
created to reproduce the angular geometry 
of the survey with constant density.
Since in the cross-correlation estimator adopted here the tracer sample is
positioned at the distance of each centre galaxy, a random catalogue does not
need to reproduce a radial selection function.
\item[(ii)]
We find that
the propagation of redshift errors onto magnitude, comoving distance and
projected distance estimates, produces systematic effects on our measurements.
We correct for these biases by modifying the projected separations involved in our
calculations using a method { tested} with the mock catalogues.
\item[(iii)]
{ We use $\omega(\sigma)$ to estimate the projected correlation function, $\Xi(\sigma)$, following
Padilla et al. (2001), Ratcliffe et al. (1998) and Croft, Dalton \& Efstathiou (1999).  Our interest
in the $\Xi(\sigma)$ correlation function lies in that it can be used to obtain the real-space
correlation function, our final objective.
The functions $\omega(\sigma)$ and $\Xi(\sigma)$ can be related via
\begin{equation}
\omega(\sigma)=B \Xi(\sigma)
\end{equation}
where the constant $B$ takes into account the selection function, $\psi$, of
the tracer sample and the individual comoving distances to the centre galaxies,
\begin{equation}
B=\frac{\sum_i \psi(y_i^p)}
{\sum_i (1/y^p_i) \int_0^{\infty} \psi(x)x^2dx}.
\end{equation}
In this equation, $y_i^p$ is the comoving distance to the $i$th centre
galaxy calculated using its redshift (spectroscopic or photometric), and the integration
variable $x$ is comoving distance.  In turn, the correlation function $\Xi(\sigma)$
bears a close relationship to the real-space correlation function $\xi(r)$ via
\begin{equation}
\Xi(\sigma)=2\int_{0}^{\infty}\xi(r=\sqrt{\sigma^2+\pi^2}) d\pi,
\label{eq:Xi}
\end{equation}
where $\pi$ is the radial component of the 3D separation $r$.  }
\item[(iv)]
For the case of approximating the real-space correlation function by 
a power law { with a slope $\gamma$}, as $\xi(r)\simeq\left(\frac{r}{r_0} \right)^{\gamma}$,
Equation \ref{eq:Xi} simplifies to
\begin{equation}
\Xi(\sigma)=r_0^{\gamma} \left[ \frac{\Gamma\left(\frac12 \right) \Gamma\left(\frac{\gamma-1}2\right)}{\Gamma\left(\frac{\gamma}2\right)}  \right] \sigma^{1-\gamma},
\label{eq:powerlaw}
\end{equation}
{ where $\Gamma(x)$ is the Gamma function.}
We use this relation to calculate the power law correlation
length, $r_0$, and slope, $\gamma$, for each subsample by minimising the quantity
\begin{equation}
\chi^2=\sum_i\frac{(\Xi^{\rm measured}(\sigma)-\Xi^{\rm fit}(\sigma))^2}{\epsilon(\sigma)^2},
\end{equation}
where the index $i$ runs over the bins in projected separation $\sigma$, { $\Xi^{\rm measured}(\sigma)$ is
the measured projected correlation function}, $\Xi^{\rm fit}(\sigma)$ is the estimate from equation \ref{eq:powerlaw},
and $\epsilon(\sigma)$ is the error in the measured correlation function calculated using the jacknife technique
(see Section \ref{ssec:recovery}).
\end{itemize}

\begin{figure}
{\epsfxsize=8.5truecm 
\epsfbox[40 170 575 705]{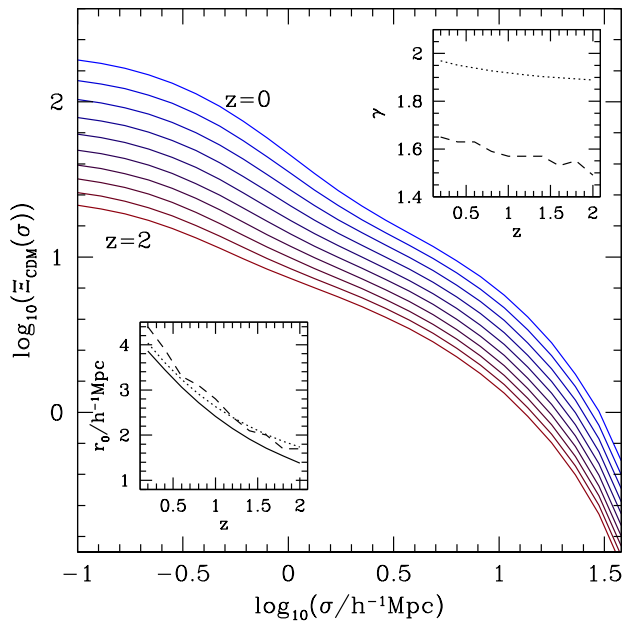}}
\caption{
Theoretical projected correlation functions for different redshifts.  The insets
show the evolution of the correlation length and power law slopes.  The solid
line shows $r_0^{\rm crossing}$, dotted lines show $r_0^{\rm rs-fit}$ and $\gamma^{\rm rs-fit}$,
and dashed lines show the resulting $r_0$ and $\gamma$ values from applying
our proposed $\chi^2$ method.
}
\label{fig:xith}
\end{figure}

{ The method outlined above is a variant of the more common procedure of
inverting the real-space correlation function from the angular correlation
function $\omega(\theta)$ (where $\theta$ is the angular separation between a pair
of galaxies) using Limber's equation (Limber, 1954).  In our case, however,
the use of $\omega(\sigma)$ introduces the use of (i) the distance to centre galaxies which in this work come
from photometric redshift estimates and (ii) the redshift distribution of
tracers which also comes from photo-zs.  This poses the question of whether photometric redshift
errors introduce important systematics in our measurements; this is answered in Section 5 where
we carry out several tests of the robustness of the method using mock catalogues.  It should
be stressed that the effect of the photo-z errors would still be similar in an
inversion of the angular correlation function using Limber's equation since in this case the photo-z errors
would affect the redshift distribution of both, centres and tracers.  Our method has the advantage of allowing
the use of different redshift ranges for tracers so as to maximise the number of pairs for
centre galaxies near the borders of a redshift bin.}

\subsection{Extracting $r_0$ from a projected correlation function, $\Xi(\sigma)$.}

In this subsection we present an attempt
to understand the process of inferring a correlation length $r_0$ and power law
slope $\gamma$
{ using} a theoretical projected cross-correlation function $\Xi(\sigma)$, paying
particular attention at the relation between the parametrisation of this power-law and
the physical quantities encoded in the correlation function.

The actual shape of the real-space correlation function deviates
from the power law proposed in Section \ref{sec:method} both in predictions
from a $\Lambda$CDM model (e.g. Zheng et al., 2005) and in observations 
(e.g. Zehavi et al., 2004).  
The meaning of $r_0$ in a power law correlation function is that of the separation
at which the correlation function satisfies $\xi(r_0)=1$.  In the case 
where the shape of $\xi(r)$ is different than a power law, we will use
the same equality to define $r_0$.
With respect to the power law slope $\gamma$, notice that 
its value will depend on the scales used to fit an estimate
of a correlation function.

We use theoretical estimates of the real-space and projected correlation functions for
the $\Lambda$CDM cosmology obtained from non-linear power spectra using the 
Smith et al. (2003) formalism.  For real-space correlations
we calculate the value of $r_0$ in { three different
ways,
(i) searching the separation,  $r_0^{\rm crossing}$, at which $\xi(r_0^{\rm crossing})=1$,
(ii) by fitting a power law to the real-space correlation function $\xi(r)$ between
separations of
$-1<\log_{10}( r/$h$^{-1}$Mpc$)<0.3$,
 in which case we obtain $r_0^{\rm rs-fit}$;
additionally, this
procedure also provides an estimate of
the power law slope, $\gamma^{\rm rs-fit}$.  
(iii)
By using the method described in the third item of the previous section (Eq. 7) of
fitting a power law to the projected correlation function between separations of 
$-1<\log_{10}( \sigma/$h$^{-1}$Mpc$)<0.3$ 
(corresponding to the scales we will use for the measured
projected correlation functions).}
The value of $r_0^{\rm crossing}$ can
be considered as the ``true" underlying value of $r_0$ which will not depend on
the parametrisation of $\xi(r)$.

Figure \ref{fig:xith} shows projected correlation functions for different
redshifts (top line for $z=0$ to bottom line for $z=2$); the inset on the lower
left shows the values of $r_0^{\rm crossing}$ as a solid line (true value), and of 
$r_0^{\rm rs-fit}$ as a dotted line; as can be seen both definitions of a 
correlation length agree to $\sim 0.2$h$^{-1}$Mpc at $z=0$ and to 
$\sim 0.4$h$^{-1}$Mpc at $z=2$.  The inset on the upper right shows as a dotted line
the resulting values of $\gamma^{\rm rs-fit}$.

The procedure outlined in item (iii) of Section \ref{sec:method} 
recovers the correlation length and power law slope, $r_0$ and $\gamma$, shown
as dashed lines in the lower left and upper right insets, respectively.  
This procedure reproduces the process that we will apply to our real data, and therefore
can be used to put into context the meaning of the measured values of $r_0$ and $\gamma$
in terms of the underlying values $r_0^{\rm crossing}$ and $r_0^{\rm rs-fit}$ and $\gamma^{\rm rs-fit}$.
As can
be seen the recovered correlation length from the projected correlation function
following the $\chi^2$ method,
is consistent with the direct fit to the real-space correlation function.  The
power law slope, on the other hand, shows a systematic offset which could be
taken into account when analysing the measured projected correlation functions.  The origin
of this offset comes from the mix of scales characterised by different correlation
function slopes, produced by
the integral over the radial direction.  In this sense, the measured value of $\gamma$ obtained
from $\Xi(\sigma)$ is a different quantity than $\gamma^{\rm rs-fit}$, the average slope of the
real-space correlation function.  In our analysis of MUSYC data { we will adopt a fixed value
for this parameter of $\gamma=-1.8$ roughly consistent with previous estimates
for galaxies at similar redshifts and also with theoretical values such as $\gamma^{\rm rs-fit}$;
the statistics only allow one parameter to be obtained from this set of galaxies.}

\section{Tests of the method}
\label{ssec:tests}

In this section we perform two separate tests of our method.  
The first is an analysis
of possible biases in the estimate of a projected distance using the photometric
(instead of spectroscopic) redshifts of centre galaxies; and the second is
a test of the recovery of the underlying clustering amplitude 
using mock catalogues, a process that takes into account the geometry of the MUSYC
fields we use, as well as possible problems due to the limited number of galaxies in
our subsamples.

\subsection{Correcting for biases in the projected separations between centre and tracer galaxies 
obtained from photometric redshift information}
\label{ssec:proj}

We now test whether the projected separations measured using
the photometric redshift of the centre galaxies are comparable to
those obtained using the spectroscopic redshifts in the mock catalogues.

\begin{figure}
{\epsfxsize=8.5truecm 
\epsfbox[40 170 575 705]{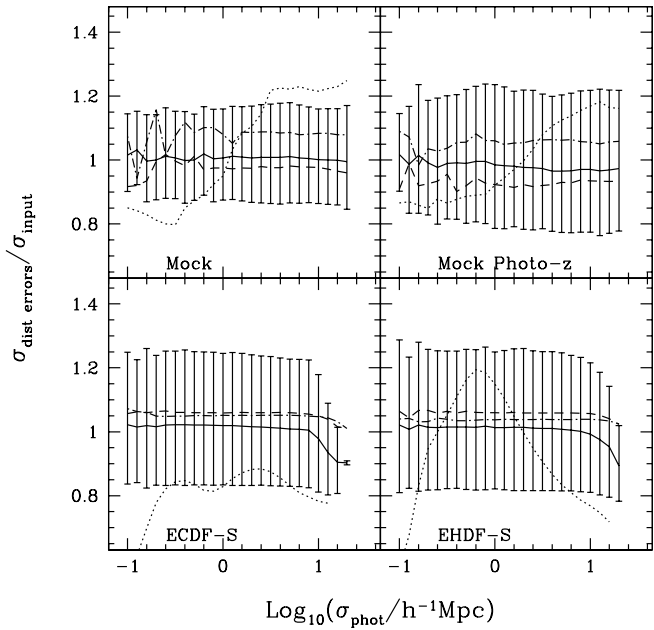}}
\caption{
Median (lines) and $10$ and $90$ percentiles (errorbars) of the distribution of
the ratio between projected separations obtained from photometric (or convolved photometric)
redshift information to real (or photometric) redshift information as a function
of projected separation,
for different redshift
bins (in dotted, solid, dashed and dot-dashed lines from the lowest to the
highest redshift subsamples); percentiles are only shown for the $0.46<z<0.68$ sample.  
Top left: results from the mock catalogues using the 
estimates of photometric redshifts and the underlying values.  Top right:
using the photometric redshifts with a further gaussian smearing corresponding
to the photo-z uncertainty from Figure \ref{fig:zphoto}
against photo-z.  The latter procedure
can be applied to real data.  Lower panels: results for the ECDF-S and EHDF-S
MUSYC fields (left and right, respectively) for photometric vs. spectroscopic redshifts where available.
}
\label{fig:cambio}
\end{figure}

We calculate projected separations using all the galaxies in
our subsamples at different redshifts to check for variations
with the distance to the observer.  In a first approach we use the
spectroscopic (true in the case of mock catalogues) and photometric redshifts.
The top-left panel of
Figure \ref{fig:cambio} shows
the ratio between photometrically and spectroscopically determined projected separations
as a function of the projected separation obtained using photometric redshifts.
Different line types correspond to different
redshift slices (always selected using the photometric redshift estimate).
The errorbars indicate the $10$ and $90$ percentiles of the distribution of ratios
in the $0.46<z<0.68$ bin.
As can be seen, the ratio shows deviations from unity, which are stronger
for the lowest redshift subsample, and only marginal for the highest redshifts probed.
The large systematic bias at lower redshifts is due to the effect of a $\Delta z\simeq 0.1$
redshift error on a distance of $z\sim0.2-0.4$, which can produce variations
in the projected separation of up to a factor of $1.5$; at larger distances
this becomes almost negligible.  Additionally, large-scale
structures can produce more important effects on the smaller volume of the lowest
redshift samples.

The measurement of the ratio between inferred and underlying projected
separations could in principle be used to correct the
projected separations obtained from the data, but would depend on the mock catalogue
selected to calculate this bias.  We devise an alternative application of this measurement
that only uses information
available from the observational data, and apply it to the same mock catalogue:  we use
the measured value of the photometric redshift error (cf. Figure \ref{fig:zphoto},
left bottom panel) in the mock, and apply an additional gaussian error of
this amplitude to the
measured photo-zs.  This produces a new version of photo-zs which we will refer to
as convolved photo-zs.  Then we calculate the projected separations of pairs using
the photo-z on the one hand, and convolved photo-z on the other, and show their ratios (convolved photo-z to photo-z) 
in the upper right panel
of Figure \ref{fig:zphoto}.  As can be seen, these ratios reproduce the results from the original
comparison between projected separations obtained from underlying (spectroscopic in the case of real data)
and photometric redshifts.  Therefore, as this process provides a good estimate
of the projected distance bias using only available data from the observations, we
can apply it to the MUSYC data.  The results for the ECDF-S and EHDF-S fields
are shown in the bottom left and right panels of the figure, respectively.  
Notice that the dispersion around the average values of these ratios are similar for both
MUSYC fields.  The slightly deeper photometry and infra-red coverage 
in the ECDF-S is responsible for the slightly lower
dispersion in the offsets with respect to EHDF-S.

\begin{figure*}
\begin{picture}(450,270)
\put(-20,-10){\psfig{file=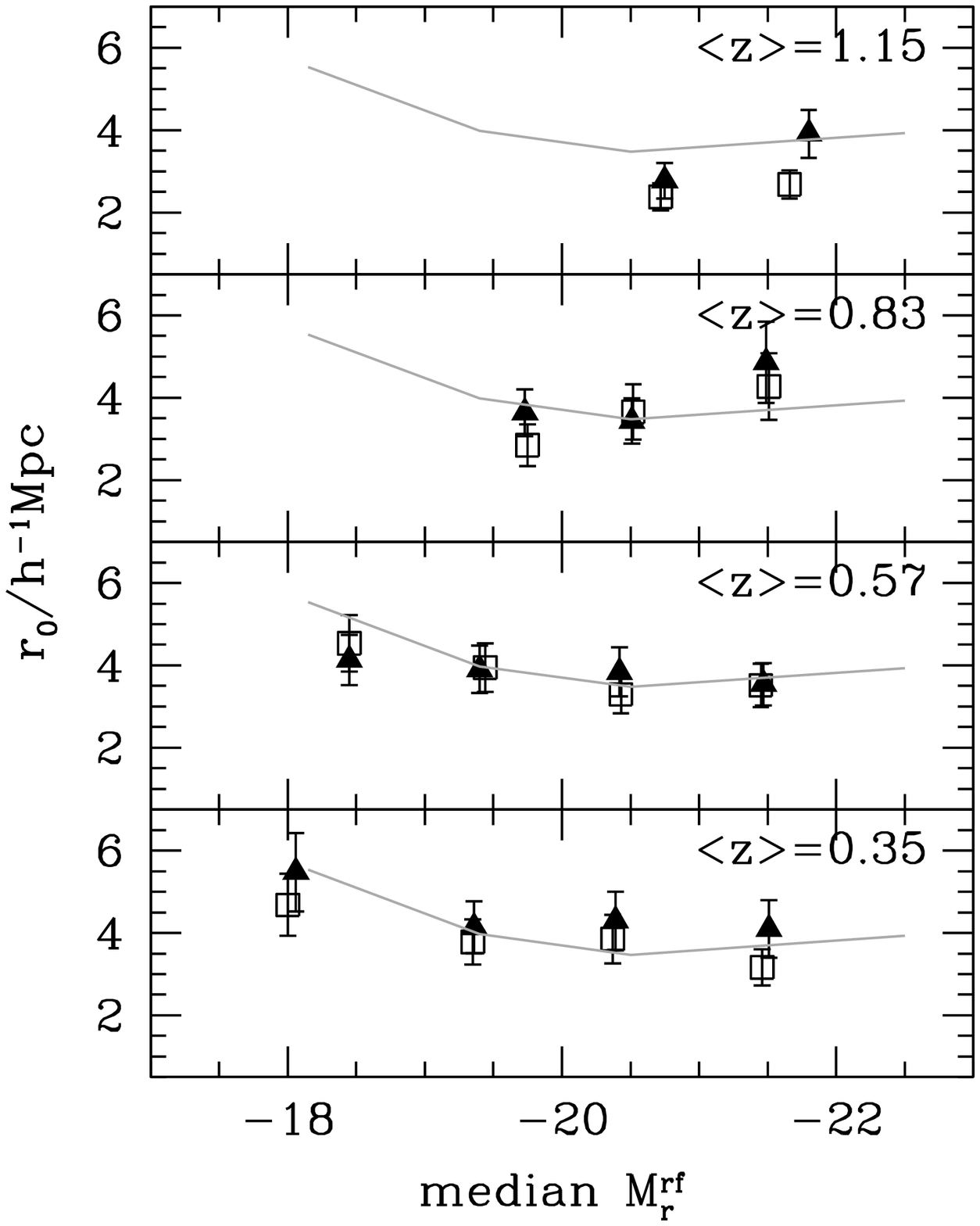,width=10.cm}}
\put(230,-10){\psfig{file=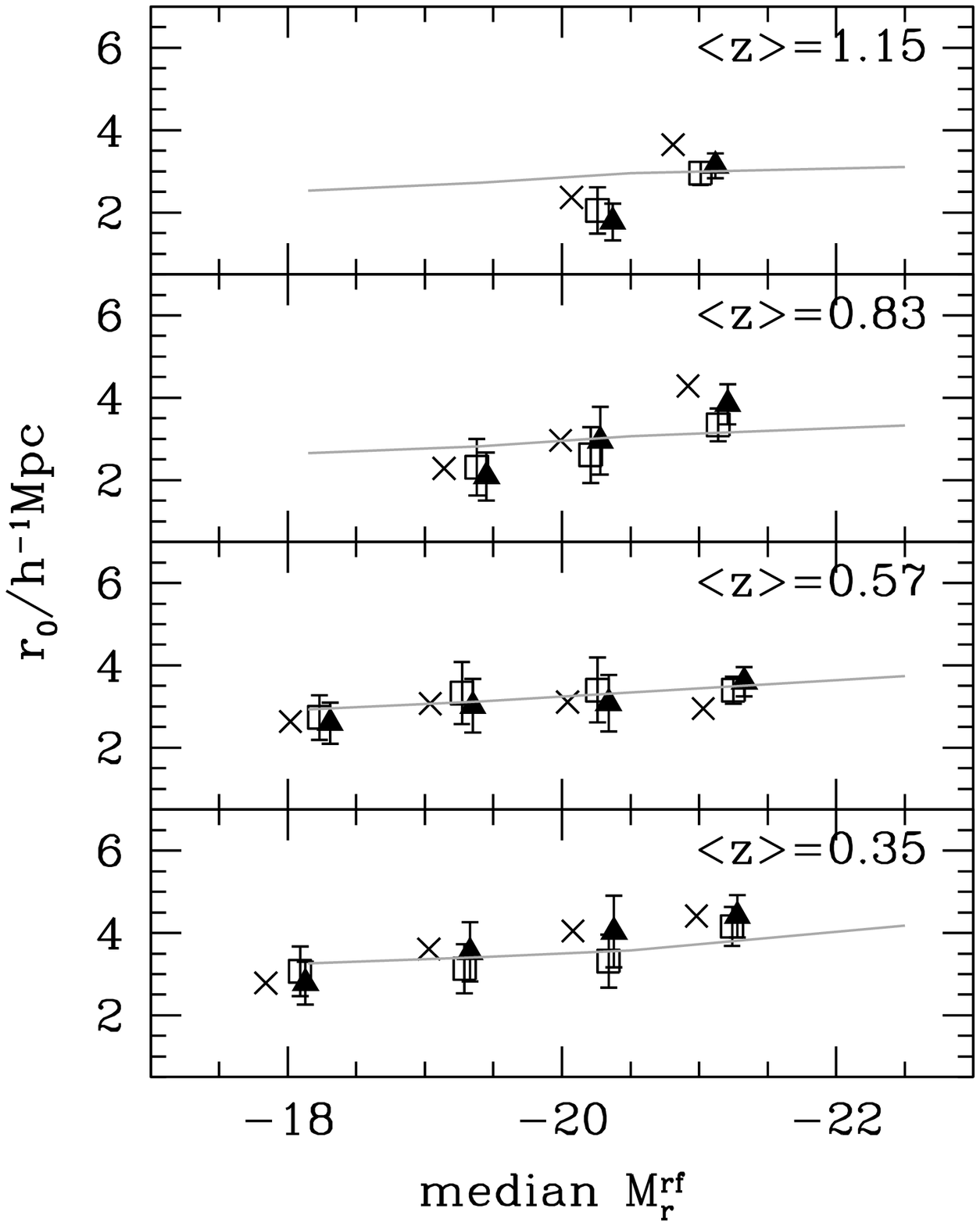,width=10.cm}}
\end{picture}
\caption{
Recovered values of $r_0$ for two mock ECDF-S catalogues (left panel for Baugh et al., 2005, galaxies,
and right panel for Bower et al., 2006, galaxies), using their
spectroscopic and photometric redshifts (open squares and filled triangles, respectively).  The right panels also show the results for the EHDF-S mock as crosses.
Different subpanels correspond to different redshift slices.  The solid
lines show the underlying clustering amplitude as a function of absolute magnitude in the simulation.
}
\label{fig:r0mock}
\end{figure*}

The corrections for the mock and MUSYC fields will be applied to the projected
separations before attempting a recovery of the spatial clustering amplitude.
As we will show in the following subsection,
this bias is the main contributor to an offset in the clustering
amplitude when using a projected correlation function and photometric redshift
information.  
Once this is taken into account, the recovery
of the clustering from the mock catalogues is similarly successful when using
either simulated spectroscopic or simulated photometric redshifts.

\subsection{Recovery of the underlying clustering amplitude in mock catalogues}
\label{ssec:recovery}

We measure the projected-angular correlation function in the mock subsamples
following the method outlined in Section \ref{sec:method} including
the correction for the
bias present in the measurement of projected separations
from using photometric
redshifts.   We use the measured correlation
functions for projected separations in the range $-1<\log_{10}( \sigma/$h$^{-1}$Mpc)$<0.3$.

When fitting the projected correlations with the power law model for $\xi(r)$,
we encounter a strong degeneracy in the likelihood of the fit as a function of $r_0$ and $\gamma$.
We lift this degeneracy by choosing a suitable value for this parameter, $\gamma=-1.8$.
As we will show in this section,
this choice provides values of $r_0$
consistent 
with the underlying values of correlation length.

{ The mock catalogue constructed using the Baugh et al. (2005) galaxies
was extracted from a single $z=0$ output of the
numerical simulation and, as a result, the underlying correlation length does not vary
with redshift.  However, the model does include a luminosity dependence
of clustering, which we calculate directly from the simulation cube.  This
is shown, as a function of absolute magnitude,
as a solid line in the left sub-panels of Figure \ref{fig:r0mock}
(the mean redshift of the samples is indicated in each sub-panel; we do not show
the median
redshift of each subsample, since these depend slightly on the
absolute magnitude limit cuts).
As can be seen, this particular model shows a decreasing correlation
length as the absolute magnitude decreases (luminosity increases) 
from $M_r=-18$ to $-20.5$, and from then on the clustering increases with
the galaxy luminosity\footnote{Measurements of this dependence
from observational data indicate a steady increase of the clustering length 
with luminosity
(i.e. Norberg et al., 2002, Zehavi et al., 2005) only when the
sample of galaxies comprises both blue and red objects.  In the
case of galaxies from the red-sequence, the dependence of clustering
is qualitatively similar to that shown by our mock catalogues
(Zehavi et al., 2005, Tegmark et al., 2004; Swanson et al., 2008,
Cresswell \& Percival, 2009).}.  The right panels show
the results for the mock with redshift evolution constructed using
Bower et al. (2006) galaxies; the solid line shows the underlying values of the clustering
length and its dependence on luminosity and redshift (these galaxies do show the higher
clustering for brighter galaxies as in Norberg et al., 2002; they also show a lower clustering amplitude 
at higher redshifts).   In all the panels, the symbols
represent the recovered values of $r_0$ from using the spectroscopic
and photometric redshifts in the volume-limited ECDF-S mock samples as solid triangles and open squares,
respectively.  The right panels also show the recovered correlation lengths for the EHDF-S mock in crosses.}
Errorbars are calculated using the jacknife technique, applied by constructing $10$ subsamples
of galaxies from a given mock catalogue (this will also apply to the analysis of the real MUSYC fields) 
by removing a different $10\%$ of its
galaxies for each jacknife subsample.  The errorbars are the dispersion with respect
to the mean in the results from each jacknife.
As can be seen, the recovery of the underlying correlation length is
quite successful, regardless of the use of photometric or spectroscopic redshifts (except for
the highest redshift samples which start to show slight differences with the underlying clustering
specially when using photometric redshifts),
indicating that to the level of certainty allowed by sets of data such
as the MUSYC fields (limited by sample variance, poisson noise), 
the bias in the projected distance is the most
important factor to take into account in this measurement.  These results also indicate that the
effect from the width of the distribution in the ratio between measured and
true projected distances does not affect the mock results to a detectable degree.
{ It is also noticeable the slightly lower performance shown by EHDF-S mock.}

\begin{figure}
\begin{picture}(200,190)
\put(20,-10){\psfig{file=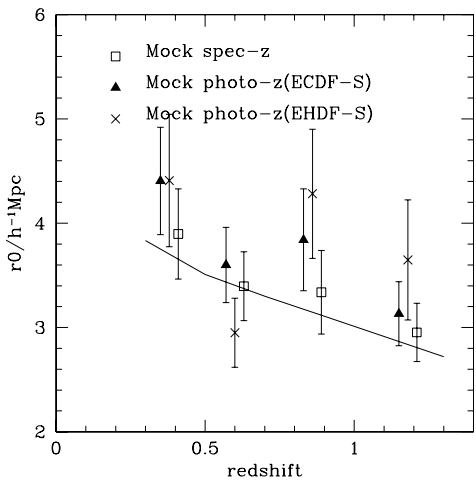,width=7.cm}}
\end{picture}
\caption{
Underlying and recovered dependence of the correlation length with redshift for the mock
catalogues constructed using Bower et al. (2006) semi-analytic galaxies; the results
are only shown for the $M_r<-21$ sample,
when adopting spectroscopic redshift in open squares, ECDF-S photometric
redshifts in solid triangles, and EHDF-S (no near infra-red photometry) photometric
redshifts in crosses.  Errorbars correspond to the jacknife errors in the recovered
real-space correlation function.
}
\label{fig:r0zmock}
\end{figure}

\begin{figure}
\begin{picture}(150,250)
\put(10,-10){\psfig{file=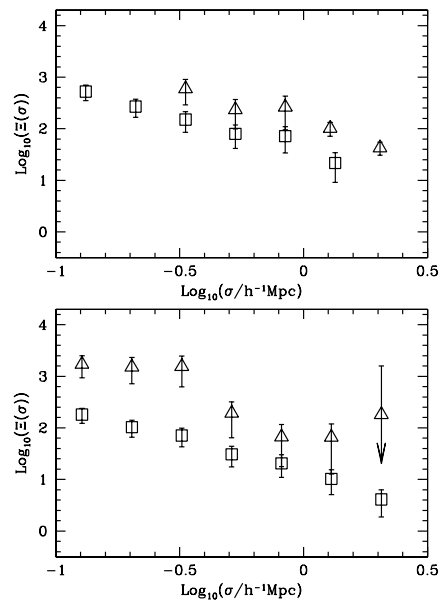,width=9.cm}}
\end{picture}
\caption{
Measured ECDF-S projected correlation functions for
galaxies 
with no restriction on template type (open squares) and for
the ETG sample of $M_r(z=0)<-19.7$.  
Top panel: galaxies
with $M_r<-21$ and $0.98<z_{\rm photo}<1.45$.
Bottom panel: galaxies
with $-20<M_r<-18.8$ and $0.1<z_{\rm photo}<0.46$.
Errorbars are calculated using the
jacknife technique.  
}
\label{fig:xiproj}
\end{figure}

Figure \ref{fig:r0mock} also shows that in the simulation, the variation of the underlying correlation
length with luminosity is $\Delta r_0\sim 1.5$h$^{-1}$Mpc between median $M_r^{rf}=-21.5$ to $-18$, for both
mocks.  { As can be seen, the use of the projected correlation function does allow to
detect the underlying dependence of clustering on luminosity with some statistical certainty 
using samples of galaxies resembling our MUSYC ECDF-S field}.  The propagation of the photo-z errors into the 
estimate of the absolute magnitude involved in the sample selection does not appear to play
an important role in this result due to the small variation of $r_0$ with intrinsic luminosity. 

We analyse the recovery of the underlying evolution of the clustering amplitude with
redshift present in the Bower et al. (2006) galaxy mocks, concentrating on the highest luminosity
subsample.  Figure \ref{fig:r0zmock} shows as a solid line the underlying dependence of the correlation length with
redshift for galaxies with $M_r<21$ in the simulation; notice that the amplitude of the variation is of
$\sim1Mpc/h$ in the range of redshifts $z=0.3$ to $1.4$.  As can be seen, the best agreement with the true
evolution is found when using spectroscopic redshifts; it can also be seen that
it is not possible to make a { highly significant} detection of such a small variation with
redshift using photometric redshift data (the ECDF-S photometry provides a slightly better match).
{ However, as the results from the mock catalogues do not show systematic differences in
the inferred evolution of clustering with redshift in comparison to the true underlying evolution, 
a detection could be made on observational data characterised by a stronger clustering evolution.
} 

\begin{figure*}
\begin{picture}(450,270)
\put(-20,-10){\psfig{file=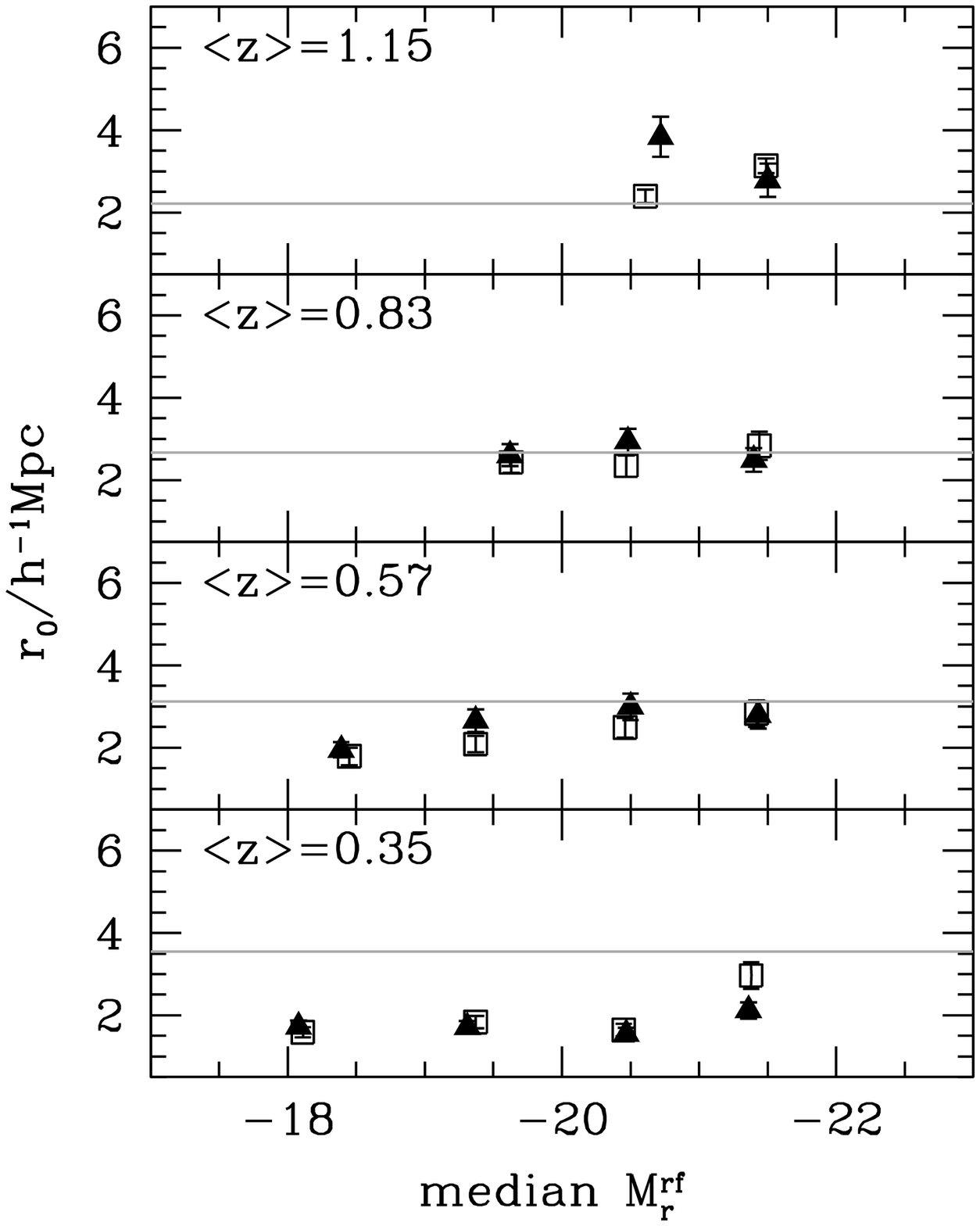,width=10.cm}}
\put(230,-10){\psfig{file=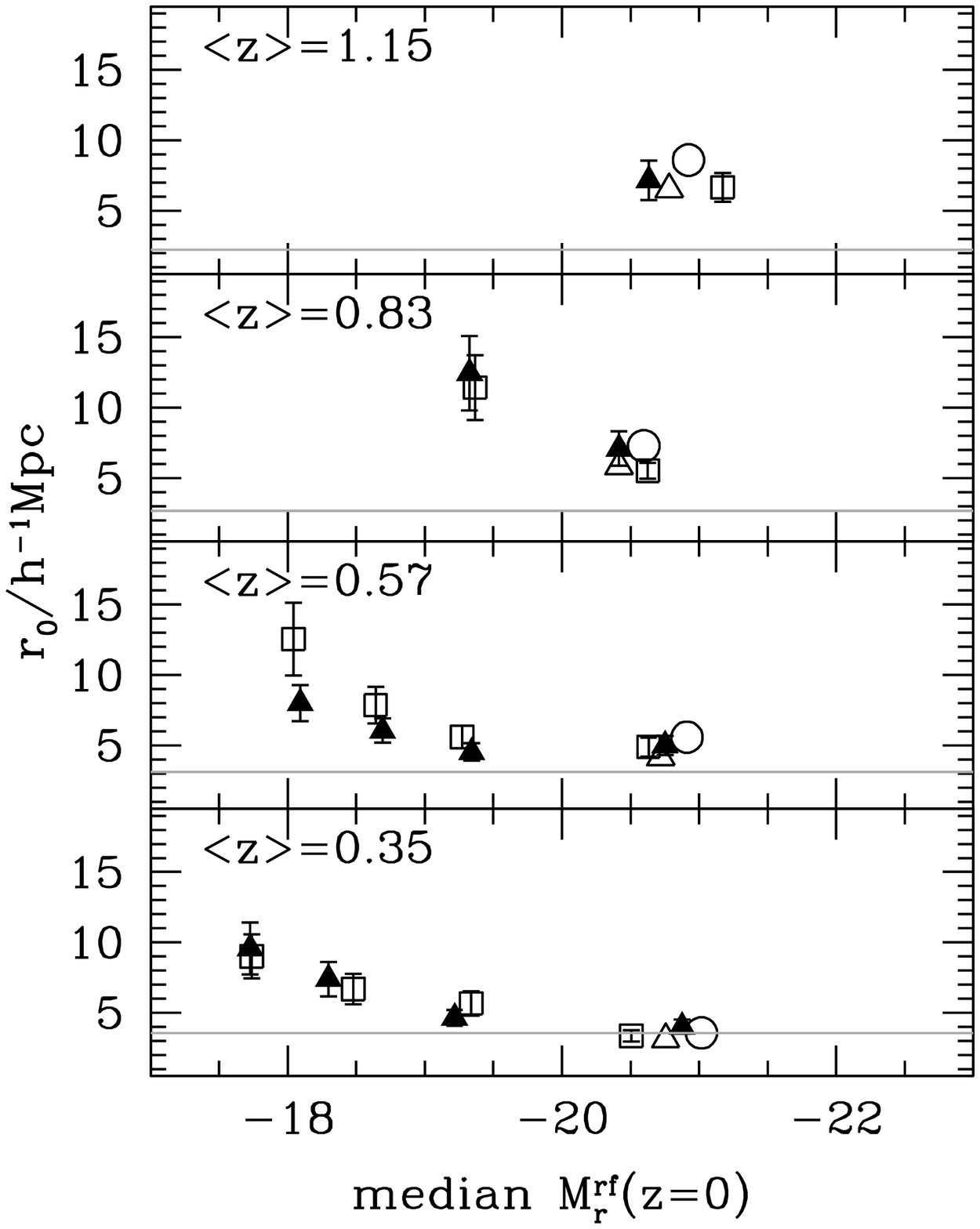,width=10.cm}}
\end{picture}
\caption{
Measured values of $r_0$ for the ECDF-S and EHDF-S fields
(filled triangles and 
open squares, respectively).  The left panel shows the result
considering all template types; the right
panel shows the results for the early type galaxies, with
similar evolved intrinsic luminosities.  The right panel also includes
the resulting clustering strength of ECDF-S samples selected by applying a
lower flux limit in the K-band (open circles) for the brightest samples at each redshift {(only
shown for the brightest samples in each redshift range, to improve clarity), and the results
for ETGs selected as red-sequence galaxies occupying the red branch of the bimodal colour 
distribution (open triangles)}.
Different subpanels correspond to different redshift slices.  The horizontal 
lines show the expected correlation length for the mass at the redshift of the sample
in a $\Lambda$CDM model.
}
\label{fig:r0data}
\end{figure*}

The lack of a perfect match between the measured (from spectroscopic redshifts) 
and underlying values of clustering
amplitude are a consequence of field-to-field variations in the simulation.  These
should also be expected in the results from the MUSYC fields at a level of 
$\Delta r_0\simeq 1$h$^{-1}$Mpc or less, affecting the lowest redshift samples (smallest volumes), 
preferentially.  For example, this variation is consistent with the different average
mass of host haloes in the Baugh et al. mock sample with $-18.8<M_r<-17.5$ and that of
all the Baugh et al. galaxies within this magnitude range in the simulation,
$<M>=1.6\times10^{13}$h$^{-1}M_{\odot}$ and  $<M>=3\times10^{13}$h$^{-1}M_{\odot}$, respectively.
The amplitude of this effect does not seem to depend on the galaxy luminosity.

We emphasise the fact that our method only uses information readily available
in the observational data, and that the photometric redshifts in our mock
catalogue are obtained following the same procedure applied to the real data.

\section{Results: The clustering of MUSYC galaxies}

We apply the method outlined in Section 4 to the real data, including the correction
of the measured projected separations using the procedure outlined
in Section \ref{ssec:proj}.  
{ Figure \ref{fig:xiproj} shows the resulting projected correlation functions 
for galaxies with no template type restriction
(squares) and ETGs (triangles) with $z>0.95$ and $M_r<-21$ (top), and $0.1<z<0.46$ and 
$-20<M_r<-18.8$ (bottom). 
We use these measurements to infer the correlation lengths, $r_0$, using Equation 7.}
The resulting dependence of
the clustering length on median rest-frame r-band absolute magnitude
and on the photometric redshift is shown in Figure \ref{fig:r0data}.
Results from the EHDF-S and ECDF-S are shown as open squares and
solid triangles, respectively;  { notice that the systematic bias expected
for the rest-frame r-band absolute magnitude obtained for EHDF-S should only
affect the highest redshift samples and shift the median magnitudes to fainter values
by up to $\sim 1$ mags.}
The horizontal grey lines show the correlation length of the mass in
a $\Lambda$CDM cosmology at the mean redshift indicated in each subpanel.
The left panel shows
the results for the subsamples extracted
from the MUSYC EHDF-S and ECDF-S fields, corresponding to different 
photo-z and $M_r$ ranges described in Section \ref{sec:data}.  
As can be seen, the results from the different fields are not entirely consistent
with one another, particularly at low redshift, { an indication that field-to-field variations are important
due to their relatively small volumes and cosmic variance}, as expected
from the analysis of the mock catalogues (differences can be larger between the
two MUSYC fields than between measured and underlying $r_0$ in the mock by 
up to a factor of $\sim 2$).
It can also be seen that up to $<z>=0.57$, there are hints at a higher clustering
for brighter galaxies.   { Most subsamples of equal luminosity cuts and specially those of ETG galaxies, 
however, show a systematic tendency to
increase their clustering with redshift, as can be seen by comparing the lowest and highest redshift cases
for each luminosity range 
(See Tables 1 and 2 for
the resulting values of $r_0$ for all the explored subsamples).
}
We will come back to this point in
the following section.  We also notice that samples corresponding to the two lowest redshift ranges
show a lower clustering than that expected for the mass (particularly for EHDF-S, and for $M_r^{rf}>-21$),
that is, a bias factor $b<1$.

\begin{table*}
\begin{minipage}{175mm}
\caption{
Resulting correlation lengths $r_0$, in units of (h$^{-1}$Mpc), 
for the EHDF-S subsamples analysed for galaxies with
no restriction on template types (top rows) and for ETGs (bottom rows).  As the redshift increases,
only the brightest luminosity samples contain galaxies.
}
  \begin{center}\begin{tabular}{@{}ccccc@{}}
  \hline
All types & & & &\\
Abs. Mags. & $0.1<z<0.46$&$0.46<z<0.68$&$0.68<z<0.95$&$0.95<z<1.45$\\
 \hline
$M_r<-21.0$       & $3.0\pm0.3$& $2.8\pm0.3$&$2.8\pm0.3$ &$3.1\pm0.2$\\
$-21.0<M_r<-20.0$ & $1.6\pm0.2$& $2.5\pm0.2$&$2.4\pm0.3$ &$2.4\pm0.2$\\
$-20.0<M_r<-18.8$ & $1.8\pm0.2$& $2.1\pm0.2$&$2.4\pm0.3$ &\\
$-18.8<M_r<-17.5$ & $1.6\pm0.1$& $1.8\pm0.2$&              &\\
  \hline
ETG gals. & & & &\\
\hline
$M_r(z=0)<-19.7$       & $3.4\pm0.4$& $4.9\pm0.7$ & $5.5\pm0.6$&$6.5\pm1.0$\\
$-19.7<M_r(z=0)<-19.2$ & $5.7\pm0.8$& $5.6\pm0.8$ & $11.4\pm2.3$ &\\
$-19.2<M_r(z=0)<-18.2$ & $6.7\pm1.1$& $7.9\pm1.3$ & &\\
$-18.2<M_r(z=0)<-17.0$ & $9.0\pm1.6$&$12.6\pm2.6$ & &\\
\hline
\label{table:1}
\end{tabular}
\end{center}
\end{minipage}
\end{table*}

\begin{table*}
\begin{minipage}{175mm}
\caption{
Same as Table 1, for the ECDF-S field.
}
  \begin{center}\begin{tabular}{@{}ccccc@{}}
  \hline
 All types & & & &\\
Abs. Mags. & $0.1<z<0.46$&$0.46<z<0.68$&$0.68<z<0.95$&$0.95<z<1.45$\\
 \hline
$M_r<-21.0$       &$2.1\pm0.2$ &$2.8\pm0.3$ &$2.5\pm0.3$ &$2.8\pm0.4$\\
$-21.0<M_r<-20.0$ &$1.6\pm0.1$ &$3.0\pm0.3$ &$2.9\pm0.3$ &$3.8\pm0.5$\\
$-20.0<M_r<-18.8$ &$1.7\pm0.2$ &$2.7\pm0.3$ &$2.6\pm0.3$ &\\
$-18.8<M_r<-17.5$ &$1.7\pm0.2$ &$2.0\pm0.2$ & &\\
  \hline
 ETG gals.  & & & &\\
\hline
$M_r(z=0)<-19.7$       &$4.0\pm0.5$ &$5.0\pm0.7$ &$7.1\pm1.2$ &$7.2\pm1.4$\\
$-19.7<M_r(z=0)<-19.2$ &$4.6\pm0.6$ &$4.5\pm0.6$ &$12.5\pm2.6$ &\\
$-19.2<M_r(z=0)<-18.2$ &$7.4\pm1.2$ &$6.1\pm0.9$ & &\\
$-18.2<M_r(z=0)<-17.0$ &$9.6\pm1.8$ &$8.0\pm1.3$ & &\\
\hline
\label{table:1}
\end{tabular}
\end{center}
\end{minipage}
\end{table*}

The right panel of Figure \ref{fig:r0data} shows the results for
the early type galaxies (notice the change of the scale in the y-axis), 
selected by a restriction to early-type galaxy templates, and a cut on rest-frame r-band
luminosity that depends on redshift so as to take into account the
aging of the stellar population of galaxies with no
significant star formation activity.  This is done using
the r-band version of the empirical model adopted by Cimatti et al.
(2006),
where the brightening of a galaxy luminosity towards higher
redshifts (or luminosity dimming as time passes) scales as $\Delta M_r=0.98 \times z$.
It can be clearly seen that we have detected a higher clustering
for early type galaxies than for the general population.
There is also some evidence for an increase of clustering with
redshift (see below) and, for the two lowest redshift bins, of a lower amplitude of clustering for brighter
objects in both fields.  The latter would
represent the first measurement of such an effect at relatively high
redshifts.  A similar dependence of clustering with luminosity has also been found for
red SDSS galaxies in the nearby Universe (Tegmark et al., 2004; 
Zehavi et al., 2005; Swanson et al., 2008). 
An analysis of the Baugh et al. semi-analytic
galaxy population
shows that this can be produced by a large number of faint satellites in high
mass DM haloes (clusters of galaxies).  This result sheds some light on
the similar trend of clustering amplitude with luminosity
shown by MUSYC and low redshift early type galaxy samples, which
could be due to a large population of intrinsically faint early type galaxies in high-mass 
concentrations.

Given the possibility that the selection of ETGs using a flux limit on $m_R$ is not complete,
we also show the resulting clustering when ETGs are extracted from samples selected in the K-band
using $m_K<22.5$ (open circles, shown only for the brightest ETG samples at each redshift {
to improve clarity; results for the other subsamples are also in agreement with those shown
by the filled triangles}).  This limit produces 
samples with similar clustering amplitudes as the selection using R-band fluxes 
showing that the detection of sources from the deep co-added $BVR$ photometry is
able to detect a reasonably complete sample of ETGs at these redshifts.  {
Additionally, 
the average observer-frame $BVR-K$ colour for our ETG templates in the redshift range $0.3<z<1.45$,
are reasonably
well described (accuracy of $15\%$) by $(BVR-K)_{ETG-of}=(1.6+2.2\times z)\pm 0.3$, which at the 
limiting redshift of our samples $z=1.45$, translates the $K=22.5$ limit into $BVR=27.3\pm0.3$, 
within the range of our detection threshold. 
}

{ We also test whether a selection of ETGs by colour instead of template types
produces significantly different results.  In order to do this we find the lower limit in observer-frame
$B-R$ colour that separates blue and red galaxies as a function of redshift.  The resulting
clustering for the red population (corresponding to the red-sequence) is shown by the open triangles
in the right-panel of Figure \ref{fig:r0data}, where as can be seen, there is a good agreement with the
results from the selection by template types.}

\section{Discussion}

\begin{figure*}
\begin{picture}(350,350)
\put(-20,-10){\psfig{file=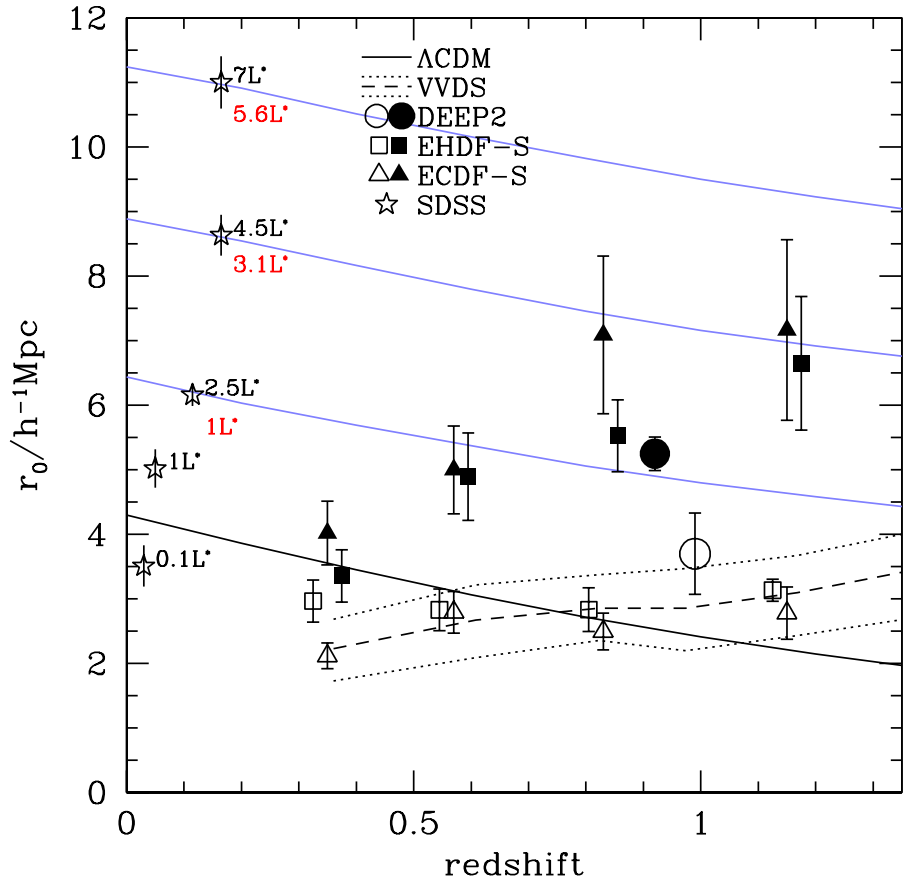,width=13cm}}
\end{picture}
\caption{
Measured values of $r_0$  
from the projected correlation function measured from bright samples
of ECDF-S and EHDF-S (triangles and squares) with no template restrictions
(open symbols) and for early types only (small solid symbols); symbols
have been slightly displaced along the x-axis to improve clarity.
{ The large open and filled circles show the results from the DEEP2 survey for
the full galaxy population (from Coil et al., 2004) and for red galaxies
(Coil et al., 2008)}; the dashed
line shows the VVDS results (the dotted lines indicate their $1-\sigma$ uncertainty)
The solid lines
show the expected evolution in a $\Lambda$CDM universe for the mass (black),
and for the descendants of DM haloes of different masses (increasing towards the
upper lines, in light blue).  
The open stars show the clustering length of SDSS galaxies of different
luminosities shown in units
of $L^*$ in black
for the full galaxy population, and in red for early-type galaxies alone.
}
\label{fig:r0z}
\end{figure*}

A global view on the evolution of the clustering of galaxies can be found
in Figure \ref{fig:r0z}, where we show the measured values of $r_0$ as 
a function of median redshift for the sample of galaxies corresponding
to the brightest absolute magnitude cut, $M_r<-21$, when
making no distinction on best-fitting template (open symbols), and
for the brightest samples of early types alone (small solid symbols).  As can be seen, early type
galaxies show a higher clustering than the samples with early and late types, and there is
a mild trend of an increasing clustering length with redshift { for the ETG samples.  The
samples with no restriction on template types are consistent with roughly constant values of
correlation length
of $r_0=(2.6\pm0.3)$h$^{-1}$Mpc for the ECDF-S field, 
and $r_0=(3.0\pm0.4)$h$^{-1}$Mpc for the EHDF-S field, throughout the redshift range considered.}
For comparison, we show as a dashed line the results from the VVDS survey (Le F\`evre et al., 2005;
the dotted lines show the errors),
and as a large open circle the results from the DEEP2 survey (Coil et al.,
2004).  In both cases the results
correspond to similar rest-frame luminosities as the MUSYC samples
with no restriction on template { shown in this figure}.  As can be seen our
results are in very good agreement with these two previous works.  { We remind the reader that }our results also
extend the clustering measurements to lower luminosity objects, and { that} we { also}
add a new measurement of the clustering of early-type galaxies with
similar passively evolved luminosities to $z=0$.  { Coil et al. (2008) also studied
the clustering of DEEP2 galaxies separated according to their rest-frame colours 
for galaxies at $z\simeq 1$.  The solid
circle shows the results for their red galaxy sample with equivalent intrinsic luminosities as
those characterising our ETG samples.  As can be seen our estimates are in agreement
with their measurement (particularly the EHDF-S result).}

The open stars in Figure \ref{fig:r0z} 
show the clustering of SDSS galaxies of different luminosities
indicated in black next to each symbol.  These clustering results are
extracted from Zehavi et al. (2005, they present results up to $L/L^*=6$), and are
extended to higher galaxy luminosities using the fit to the variation of the
bias factor by Tegmark et al. (2004).  The red labels show the luminosity 
of early-type galaxies (selected from the red sequence in a colour-magnitude diagram) 
corresponding to the same clustering amplitude,
as measured by Swanson et al. (2008).

The black solid line in
Figure \ref{fig:r0z}
corresponds to the evolution of $r_0$ from the smooth DM density field (calculated
using non-linear power spectra from Smith et al., 2003), and the light blue lines
show the evolution of the clustering of DM haloes of a given mass (increasing
from the lower to the upper lines)
as their evolution is followed to $z=0$ using the merger trees in the numerical 
simulation\footnote{This represents
the trend of the clustering of descendants; notice
that the variation in $r_0$ for the descendants 
is similar to that of the smooth mass density field. In a plot of the
evolution of the bias factor there is
a more clear difference between the trends shown by descendants and
by the mass;  for the latter b=1 for all redshifts.}.  
We use these lines to interpret our clustering measurements.
If the results corresponding to galaxies characterised by similar properties (intrinsic
luminosity, spectral template) are found to lie on a particular solid light blue line,
it could be considered that the lower redshift samples are direct descendants of their
higher redshift counterparts.  
As can be seen, { some} of our MUSYC subsamples
{ might not} be connected in this simple way.
In particular, the samples with no restriction on templates, which
present narrow errorbars, { could} allow a statistical refutation of perfect connection
of $\sim 5\sigma$ for ECDF-S,
$\sim 3\sigma$ for EHDF-S\footnote{This of course would still allow some galaxies in one subsample
to be descended from those in another.}.  
Luminous galaxies at $z\simeq 1.15$ { could} evolve into objects
with higher clustering than galaxies of similar rest-frame luminosity at $z\simeq 0.37$.
The present-day descendants of the bright, volume-limited ECDF-S and EHDF-S subsamples shown
here { would} roughly { be} within $0.5<L/L^*<1.5$.  {
However, the sample of galaxies with no template restriction is not particularly suitable
for this analysis since the descendants of many of the galaxies in the high-z samples 
will probably be below the lower luminosity limit of the low-z samples since we are not
taking into account any possible evolution in this case.  The selection of ETGs which
we analyse next does take into account passive evolution and we are therefore able to
analyse their descendants.
}

\begin{figure}
\begin{picture}(250,240)
\put(0,0){\psfig{file=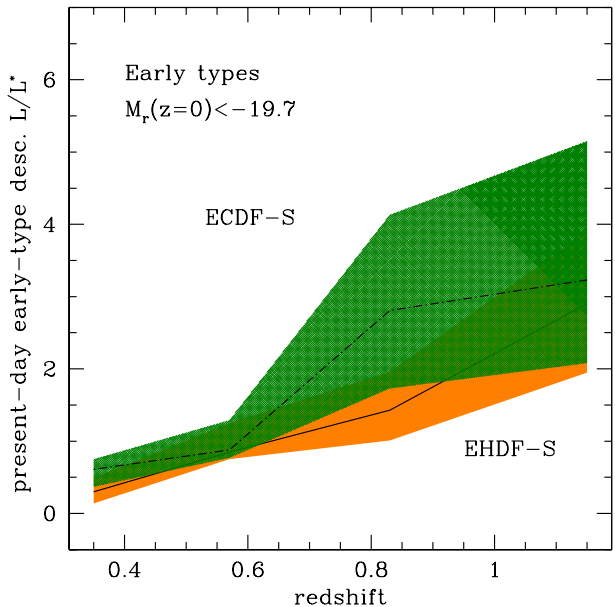,width=8.6cm}}
\end{picture}
\caption{
Typical present-day descendant luminosity $(L/L^*)$ for samples of early type
galaxies with evolved luminosities $M_r(z=0)<-19.7$, for the EHDF-S
and ECDF-S fields (solid and dot-dashed lines, respectively).  The shaded
areas indicate the uncertainty in the descendant luminosity arising from
the uncertainties in the clustering length measurements.  Descendants
are assumed to be early type galaxies.
}
\label{fig:des}
\end{figure}

For the ETG samples, the refutation of a direct descendant/parent
relationship between the $z\simeq 1.15$ and $z\simeq 0.37$ samples is of $2\sigma$ and $2.6\sigma$
significance for the ECDF-S and EHDF-S, respectively.  If this were confirmed with larger galaxy samples, it would
conflict with recent exercises where early type galaxies
with equivalent evolved $z=0$ luminosities { selected by requiring them to populate
the red sequence} are compared as descendants/parents
of each other.  Such attempts have been widely used to study the epoch of assembly of
stellar mass in early types (i.e. Cimatti et al., 2006).
Given the low SF activity characterising these samples of galaxies, their present-day
descendants should also correspond to early type galaxies (unless new SF episodes
were triggered).  Assuming that this is the case and using the Swanson et al. { SDSS} 
$z=0$ early-type galaxy luminosities (indicated in red labels), we calculate
the resulting descendant luminosity of early type
galaxies at different redshifts { by following the halo descendant tracks that connect
the ETG clustering amplitude at redshift $z$ down to $z=0$ (cf. Figure \ref{fig:r0z}), where we interpolate to find the luminosity
of the SDSS ETGs which would be characterised by this clustering amplitude.  These results are }shown in Figure \ref{fig:des}; shaded areas 
show the uncertainties calculated by propagating the errors in $r_0$.
As can be seen, the descendant 
luminosity drops from $L/L* \sim 2-5.2$ to as low as $L/L^*\sim 0.1-0.8$
from $z=1.15$ to $z=0.37$ (the ranges in $L/L^*$ encompass the results obtained
from the EHDF-S and ECDF-S fields).  We remind the reader that this result is dependent
on the rates of mergers used to trace the descendants of high-redshift galaxy samples; therefore, 
this conclusion corresponds only to the cosmological model adopted in this paper.  
We will use these results to study the assembly
of massive galaxies in  MUSYC in a forthcoming paper (Padilla et al., 2010).

\section{Conclusions}

We presented a measurement of the evolution of galaxy clustering
with redshift from the MUSYC survey.  
We used $\sim50,000$ galaxies in the MUSYC EHDF-S and ECDF-S
fields for which photometric
redshifts were calculated using a least squares frequentist best-fit method in combination
with the synthetic spectra used by the HYPERZ code along with 
a specially designed template set from Christlein et al. (2009).  
We divide the sample of galaxies into bins in photo-z delimited by
$0.1-0.46$, $0.46-0.68$, $0.68-0.98$ and 
$0.98-1.45$;
which results in $4$ subsamples
with a total number of galaxies ranging between $\sim9000-17000$ each.  
These are then further divided according to their
rest-frame luminosities.

We use a method involving the projected-angular correlation function, which
is thoroughly tested using theoretical estimates of the
projected correlation function, as well as two mock MUSYC fields.  We 
demonstrate that
the application of this technique to samples with
photometric redshift information can provide reliable results
on the clustering amplitude of galaxies, out to $z\simeq 1.5$.  

We find important systematic biases in the determination of projected
separations when photometric redshifts are used as an indicator of
the distance to a sample of centre galaxies; this effect 
arises from the important variations introduced in the distance
to galaxies by the redshift error, which
translates into a change in the projected separation.  This bias
can be
particularly important for low redshift samples, since the relative
redshift errors are larger and the small sampled volumes are more sensitive to
large-scale structure variations.  We propose and test a method
to estimate these biases, using only information available from observational
data.  We find that this is the most important bias affecting the clustering
measurement from this method, at least to the degree of certainty allowed by
the size of our samples.

The results from MUSYC galaxies indicate that the real-space correlation
length $r_0$ of $M_r<-21$ (rest-frame) galaxies { is consistent with constant
values over the redshift range explored, $0.1<z<1.45$, of
$r_0=(2.6\pm0.3)/h^{-1}Mpc$ and $(3.0\pm0.4)/h^{-1}Mpc$ 
for the ECDF-S and EHDF-S fields, respectively.}
These values are consistent within the errorbars with previous estimates from
the VVDS survey (Le F\`evre et al. 2005) and DEEP2 (Coil et al., 2004) 
for samples with similar intrinsic luminosities.  By extension, these
measurements would also be consistent with the
results for the zCOSMOS survey by Meneux et al. (2009) who obtain
similar results to the VVDS; a more direct comparison with our measurements
would involve replicating their sample selection which we do not attempt at this time.

We also studied the clustering properties of early type galaxies with similar
evolved intrinsic luminosities (using a passive evolution recipe), { finding
good agreement with a previous measurement of equivalent samples of red galaxies
by Coil et al. (2008) at $z\simeq 1$}.  Samples
selected this way at different redshifts have been proposed as tools to study
the evolution of the stellar content of early type galaxies with redshift, to infer
the typical epoch of assembly.  We find indications that such samples may
not constitute a single evolving population; furthermore, our results
indicate that $z\sim 1$ early type galaxies evolve into present-day objects
with a higher clustering than their counterparts of similar
evolved luminosity at lower redshifts. 

Our results have suffered from considerable cosmic variance {(similar to the effects
suffered by DEEP2 and VVDS)}, an issue that will
soon be overcome by larger surveys.  In particular, 
the method presented in this paper will be extremely useful to analyse upcoming or planned
photometric surveys which will cover areas many orders of magnitude larger than MUSYC.  Examples are
the Dark Energy Survey ($4,000$sq. degrees, Huan et al., 2009), 
or the Large Synoptic Survey Telescope ($20,000$ sq. degrees, Ivezic et al., 2008)
which, while
characterised by photometric redshift errors comparable to those from MUSYC,
will allow extremely accurate measurements of the clustering dependence on
redshift, luminosity and template type, and in turn on the parent-descendant relation
between samples of galaxies at different redshifts.

\section*{Acknowledgments}
We acknowledge constructive comments from an anonymous referee.
NDP was supported by a Proyecto Fondecyt Regular
no. 1071006.  This work was supported in part by the ``Centro de Astrof\'\i sica
FONDAP" $15010003$, and by BASAL-CATA.
This material is based upon work supported by the National Science Foundation
under Grant. No. AST-0807570.
We are indebted to the Durham group for kindly providing us with GALFORM 
simulation outputs.

\end{document}